\documentclass[%
 reprint,
 amsmath,amssymb,
 aps,
pre,
]{revtex4-2}

\usepackage{graphicx}
\usepackage{bm}
\usepackage{xcolor}
\usepackage{amsmath}
\usepackage{booktabs}
\usepackage{dcolumn}
\usepackage[caption=false]{subfig}
\usepackage{hyperref}

\begin{document}

\preprint{APS/123-QED}

\title{Inverse energy transfer in decaying MHD turbulence: A shell-to-shell analysis}

\author{Lenard Kasselmann}
\affiliation{Hamburger Sternwarte, University of Hamburg, Gojenbergsweg 112, 21029 Hamburg, Germany}\email{lenard.kasselmann@uni-hamburg.de}  

\author{Philipp Grete}
\affiliation{Hamburger Sternwarte, University of Hamburg, Gojenbergsweg 112, 21029 Hamburg, Germany}

\author{Pranjal Trivedi}
\affiliation{Hamburger Sternwarte, University of Hamburg, Gojenbergsweg 112, 21029 Hamburg, Germany}

\author{Marcus Brüggen}
\affiliation{Hamburger Sternwarte, University of Hamburg, Gojenbergsweg 112, 21029 Hamburg, Germany}

\author{Robi Banerjee}
\affiliation{Hamburger Sternwarte, University of Hamburg, Gojenbergsweg 112, 21029 Hamburg, Germany}

\begin{abstract}

\texttt{Context.} In decaying magnetohydrodynamic (MHD) turbulence, energy can be transported from small to large scales, known as inverse transfer. Historically, this was understood as a consequence of the conservation of magnetic helicity, a measure of the topology of magnetic field lines. However, using numerical simulations, inverse transfer was even found in systems where helicity is vanishing on average. 

\texttt{Methods.} We explore the mechanism behind inverse transfer by measuring shell-to-shell energy and helicity transfer functions from numerical simulations of non-helical and maximally helical decaying MHD turbulence. 

\texttt{Results.} Independent of magnetic net-helicity, large magnetic scales receive energy directly from the integral scale in both the magnetic and kinetic reservoirs, leading to increasingly non-local transfer for larger receiving scales. The resulting rate of energy increase in each receiving scale is proportional to its energy, resulting in self-similar, multiplicative growth. Even though the system is magnetically dominated, contributions from kinetic-magnetic and magnetic-magnetic energy-exchange are similar in magnitude. In the case of vanishing net-helicity, transfer functions between the positively and negatively helical parts of the field are computed. We find that inverse transfer only occurs within each helical sector, not across them.

\texttt{Conclusions.} Our findings are broadly consistent with the theory underlying the conservation of the Hosking integral, which explains inverse transfer as merging of local magnetic islands with equal-signed helicity.

\end{abstract}

\maketitle


\section{Introduction} 

Decaying magnetohydrodynamic (MHD) turbulence has received considerable attention in recent years due to its crucial applications in cosmology and astrophysics \cite{2022JPlPh..88e1501S}. In cosmology, it is relevant for the evolution of primordial magnetic fields (PMF) \cite{PhysRevD.70.123003, Durrer:2013pga, Subramanian_2016}. In astrophysics, it is used to describe the dynamics of the solar wind \cite{10.1111/j.1365-2966.2011.18933.x}. 

In three-dimensional hydrodynamic turbulence, as originally formalized by \citet{1941DoSSR..30..301K}, energy is typically transported from large to small scales through scale-local interactions, a process known as a direct cascade. In MHD turbulence, however, additional conserved quantities can influence the dynamics. One of these is \textit{magnetic helicity}, which measures the topology (i.e. linkage and self-twist) of magnetic field lines and is calculated as
\begin{equation}\label{eq:helicity_def}
H = \int_V \mathbf{A}(\mathbf{x})\cdot \mathbf{B}(\mathbf{x}) \:dV,
\end{equation}
where $\mathbf{B}$ is the magnetic field and $\mathbf{A}$ is its vector potential. This quantity is conserved in ideal MHD, i.e. for vanishing magnetic resistivity \cite{doi:10.1073/pnas.44.6.489}, but still approximately conserved if the magnetic Reynolds number is large enough. 
It is well known that strong magnetic helicity, i.e. $H\sim E\cdot L$ where $E$ is the total magnetic energy and $L$ is the typical scale of magnetic eddies, can cause \textit{inverse energy transfer}, i.e. transfer of energy from small to large scales, in a decaying magnetic field \citep{PhysRevD.81.123002, Pouquet_Frisch_Léorat_1976, PhysRevE.64.056405}. This is explained in terms of a selective decay argument: If helicity is better conserved than energy, then $H \sim E \cdot L$ implies that, as $E$ decays, $L$ will grow. An alternative (somewhat more heuristic way) to understand inverse transfer in the case of maximally helical fields, is to imagine the magnetic field as a sea of localized magnetic structures or islands that merge into larger structures via reconnection. 

For the sake of completeness, we note that inverse transfer can also occur in driven systems, e.g. \citet{PhysRevE.85.015302}. 

More recently, it was found using numerical simulations that inverse transfer is possible even if helicity is absent on average, albeit weaker than in the fully helical case for a Batchelor spectrum (i.e. a magnetic energy power spectrum with a subinertial range that increases with wave number as $k^4$) \citep{PhysRevLett.114.075001}. Since helicity is a signed quantity, a system with globally vanishing helicity can still contain patches of locally non-zero helicity density, see lower-right panel in Fig. \ref{fig:Helicity_density}. 
\citet{PhysRevX.11.041005} (hereafter referenced to as HS) used this fact to explain inverse transfer in systems where helicity is vanishing on average: Locally helical structures are stable, while locally non-helical structures are unstable. If two structures of equally-signed helicity merge, they form a larger structure, resulting in inverse transfer of energy and helicity. If two structures of oppositely-signed helicity merge, their helicities cancel out, resulting in a non-helical structures which decays away quickly. In other words, structures of opposite helicity annihilate.  

In this scenario, the selective-decay argument still applies, except the invariant that is driving the inverse transfer is now the \textit{Hosking integral}, a measure of large-scale helicity fluctuations. 

The theory makes predictions about the scaling of certain global quantities, such as the integral scale and magnetic energy as a function of time, as well as the evolution of the energy-spectra, which have been tested \citep{2023JPlPh..89f9006B} \footnote{For spectra with a shallower subinertial range, the dynamics may instead be controlled by other conserved quantities, such as the Saffman integral in the $k^2$ case \cite{2023JPlPh..89f9006B}.}. 

A more detailed picture of the underlying turbulent dynamics can be obtained by analyzing the nonlinear terms in the MHD equations in terms of triad interactions in Fourier space, where energy or helicity at one scale is transferred to another via interactions mediated by a third scale. In incompressible MHD, there are two energy reservoirs, kinetic and magnetic, that can exchange energy. Furthermore, decomposing the magnetic field into positively and negatively helical sectors \citep{Waleffe1992} allows one to directly track interactions between equally- and oppositely-signed helical structures.

Transfer functions in the helical mode decomposition have been studied by \citet{Linkmann_Berera_McKay_Jäger_2016} in a purely analytical way, but have so far not been measured and analyzed directly for a decaying turbulence simulation. 

In this paper, we address this research gap by applying the shell-to-shell transfer formalism \citep{PhysRevE.72.046301, DAR2001207, 2011PhPl...18k2307T, 10.1063/1.4990613} to high-resolution simulations of decaying helical and non-helical MHD turbulence. This approach allows us to study in detail how energy is exchanged across scales in the kinetic and magnetic reservoirs and between the positively and negatively helical sectors of the magnetic field in the case of vanishing net helicity. In addition, we compute shell-to-shell transfer functions for magnetic helicity in order to investigate the relation between helicity transfer and energy transfer. These diagnostics allow us to assess the role of different scale interactions in mediating inverse transfer and directly test the dynamical picture of interacting helical structures underlying HS theory. 

The rest of the paper is structured as follows: in section \ref{sec:definitions}, we define the relevant quantities, such as magnetic helicity and the Hosking integral. In section \ref{sec:numerical_methods}, we present the details of our simulations and the initial conditions. In section \ref{sec:shell-to-shell}, we review the shell-to-shell energy transfer formalism. In section \ref{sec:results}, we present the results from our simulations and their analysis. In section \ref{sec:discussion}, we discuss our results and end with a conclusion in section \ref{sec:conclusions}. Additional information is given in the Appendix.  

\section{Definitions}\label{sec:definitions}
In this section, we define the quantities which are relevant in the context of decaying MHD turbulence. Throughout the paper, bold symbols denote vector quantities. Fields written as $\mathbf{F}(\mathbf{x})$ refer to real space, while $\mathbf{F}(\mathbf{k})$ denotes their Fourier transform. Different measurements of helicity are summarized in Tab. \ref{tab:helicity}.

\subsection{Power spectra}

We define the \textit{magnetic energy power-spectrum} as 
\begin{equation}E_M(k) = \frac{1}{2} \left< \left| \mathbf{B}(\mathbf{k}) \right|^2 \right>,\end{equation}
where $\mathbf{B}(\mathbf{k})$ is the magnetic field in Fourier-space and $\left<.\right>$ denotes angle-averaging in Fourier-space. The magnetic energy power-spectrum is related to the total magnetic energy as 
\begin{equation}E_M = \int dk \: E_M(k)\end{equation}
and therefore has units of energy $\times$ length. Similarly, the kinetic energy spectrum is defined as,
\begin{equation}
E_K(k) = \frac{1}{2} \left< \left| \mathbf{U}(\mathbf{k}) \right|^2 \right>,
\end{equation}
where $\mathbf{U}$ is the velocity field. 
The magnetic helicity power-spectrum is given by 
\begin{equation}\mathcal{H}(k) = \left< \mathbf{A}(\mathbf{k})\cdot\mathbf{B}^*(\mathbf{k}) \right>,\end{equation}
such that it can be related to the total magnetic helicity as 
\begin{equation}H = \int dk \: \mathcal{H}(k).\end{equation}
From these definitions follows the realizability condition
\begin{equation}\label{eq:realizability}
    |\mathcal{H}(k)| \leq \frac{2 E_M(k)}{k},
\end{equation}
which implies that the \textit{fractional helicity spectrum}
\begin{equation}\label{eq:frac_helicity}
    h(k) = \frac{k \:\mathcal{H}(k)}{2 E_M(k)}
\end{equation}
can have values $-1\leq h(k) \leq1$. In this paper, we will study the edge-cases $h(k) = 0$ and $h(k) = 1$ for all $k$, which we will call "non-helical" and "helical", respectively. Note that in an isotropic system the sign of the total helicity does not matter and $h = +1$ and $h = -1$ will give rise to the same dynamics. 

\subsection{Helical mode decomposition}

Any three-dimensional field in Fourier space $\mathbf{F}(\mathbf{k})$ can be expressed as 
\begin{equation}
\mathbf{F}(\mathbf{k}) = f(\mathbf{k})^+\: \mathbf{e}^+(\mathbf{k}) + f(\mathbf{k})^- \:\mathbf{e}^-(\mathbf{k}) + f^\parallel(\mathbf{k}) \: \mathbf{e}^\parallel(\mathbf{k}),
\end{equation}
where $\mathbf{e}^\parallel(\mathbf{k})$ is a unit vector parallel to $\mathbf{k}$, and $\mathbf{e}^\pm(\mathbf{k})$ are unit eigenvectors of the curl operator: 
\begin{equation}
i \mathbf{k} \times \mathbf{e}(\mathbf{k})^\pm = \pm k \: \mathbf{e}(\mathbf{k})^\pm
\end{equation}
Since the magnetic field is solenoidal, its longitudinal component vanishes,
\begin{equation}
\mathbf{B}(\mathbf{k}) = b(\mathbf{k})^+\: \mathbf{e}^+(\mathbf{k}) + b(\mathbf{k})^- \:\mathbf{e}^-(\mathbf{k}).
\end{equation}
We will subsequently use the shorthand notation
\begin{equation}\label{eq:helical_decomp}
    \mathbf{B}(\mathbf{k}) = \mathbf{B}^+(\mathbf{k}) + \mathbf{B}^-(\mathbf{k})
\end{equation}
with $\mathbf{B}(\mathbf{k})^\pm = b^\pm(\mathbf{k})\:\mathbf{e}^\pm(\mathbf{k})$.

In this basis, the energy and helicity of an individual helical Fourier mode are respectively given by, 
\begin{equation}
    E_M^\pm(\mathbf{k}) = \frac{1}{2} \left| \mathbf{B}^\pm(\mathbf{k}) \right|^2
\end{equation}
\begin{equation}
    \mathcal{H}^\pm(\mathbf{k}) = \pm \frac{1}{2k} \left| \mathbf{B}^\pm(\mathbf{k}) \right|^2.
\end{equation}
Energy and helicity of an individual helical mode are thus directly related:
\begin{equation}
\mathcal{H}^\pm(\mathbf{k}) = \pm \frac{1}{k} E^\pm(\mathbf{k})
\end{equation}
Consequently, the corresponding spectral quantities satisfy
\begin{equation}\label{eq:signed_helicity_energy_relation}
\mathcal{H}^\pm(k) = \pm \frac{1}{k} E^\pm(k).
\end{equation}

\subsection{Hosking Integral}

Even if the total helicity of the system as defined in equation \ref{eq:helicity_def} is vanishing, the \textit{Helicity-density} $\mathcal{H}(\mathbf{x}) = \mathbf{A}(\mathbf{x})\cdot \mathbf{B}(\mathbf{x})$ can still be non-zero locally. In this case, one expects that patches of positive and negative helicity-density are equally abundant, such that they average to zero. The typical size of such a patch should at any given time be roughly equal to the integral scale of the magnetic field. HS proposed that these patches can merge via reconnection in such a way that the large-scale mean-square of the helicity
\begin{equation}
    I_H = \lim_{V\rightarrow \infty} \frac{1}{V} \left[\int d V \: \mathcal{H}(\mathbf{x})\right]^2,
\end{equation}
now known as the \textit{Hosking integral}, is approximately conserved.

\subsection{Gauge freedom and helicity}

While total helicity, its spectral density, and the helicity transfer functions are gauge-independent in a periodic domain, gauge terms appearing only as vanishing surface terms, the helicity density in real space is gauge-dependent. The gauge-independence of the Hosking integral is therefore more subtle, as discussed in \citet{hew_hosking_federrath_beattie_kriel_2026}. Following \citet{Subramanian_2006}, we work in the Coulomb gauge throughout, in which helicity density is directly related to the topological linking number of field lines, making the interpretation in terms of locally helical structures physically meaningful.

\newcolumntype{L}[1]{>{\raggedright\arraybackslash}p{#1}}
\newcolumntype{C}[1]{>{\centering\arraybackslash}p{#1}}
\newcolumntype{Y}{>{\raggedright\arraybackslash}X}

\begin{table*}
\centering
\begin{tabular}{lcl}
\toprule
\textbf{Type} & \textbf{Mathematical Definition} & \textbf{Description} \\
\midrule
\textbf{Total Magnetic Helicity}
& $H = \int_V \mathbf{A}\cdot\mathbf{B}\, dV$
& Measures the global linkage, twist, and writhe of \\
& & magnetic field lines within a volume. Conserved in ideal MHD. \\
\textbf{Helicity Density}
& $\mathcal{H}(\mathbf{x}) = \mathbf{A}(\mathbf{x})\cdot\mathbf{B}(\mathbf{x})$
& Local measure of magnetic helicity at each point in space.\\
\textbf{Helicity Power Spectrum}
& $\mathcal{H}(k) = \langle \mathbf{A}(\mathbf{k}) \cdot \mathbf{B}^*(\mathbf{k}) \rangle$
& Describes how helicity is distributed across spatial scales. \\
\textbf{Fractional Helicity Spectrum}
& $h(k) = \dfrac{k\,\mathcal{H}(k)}{2E_M(k)}$
& Dimensionless measure of helicity relative to magnetic \\
& & energy at scale $k$. Bounded by $-1 \le h(k) \le 1$. \\
\textbf{Signed Helicity Power Spectrum}
& $\mathcal{H}^{\pm}(k) = \pm \frac{1}{k} \langle |\mathbf{B}^{\pm}(\mathbf{k})|^2\rangle$
& Helicity spectrum for the positively and negatively \\
& & helical parts in the helical-mode decomposition. \\
\bottomrule
\end{tabular}
\caption{Different measurements of magnetic helicity.}
\label{tab:helicity}
\end{table*}

\section{Numerical methods}\label{sec:numerical_methods}
For our simulations, we use the open source, finite-volume MHD code \texttt{AthenaPK} \footnote{\texttt{AthenaPK} is available and maintained at https://github.com/parthenon-hpc-lab/athenapk and commit ca99dd4 was used for the simulations.}, which is built on top of the performance-portable adaptive-mesh refinement framework \texttt{Parthenon} \citep{parthenon} to solve the equations of compressible MHD:

\begin{align}\label{eq:MHD}
\frac{\partial \rho}{\partial t}
+ \nabla \cdot (\rho \mathbf{U}) = 0, \\
\frac{\partial \mathbf{U}}{\partial t}
+ (\mathbf{U}\cdot\nabla)\mathbf{u}
= -\frac{1}{\rho}\nabla P_{\mathrm{tot}}
+ \frac{1}{\rho}(\mathbf{B}\cdot\nabla)\mathbf{B}, \\
\frac{\partial e_{\mathrm{tot}}}{\partial t}
+ \nabla \cdot \left[
\mathbf{U}(e_{\mathrm{tot}} + P_{\mathrm{tot}})
- (\mathbf{U}\cdot\mathbf{B})\mathbf{B}
\right] = 0, \\
\frac{\partial \mathbf{B}}{\partial t}
= \nabla\times(\mathbf{U}\times\mathbf{B}),
\end{align}

where $\rho$ is the density, $\mathbf{U}$ is the velocity field, $P_{\textnormal{tot}} = P_\textnormal{gas} + \frac{1}{2}|\mathbf{B}|^2$ is the total pressure, $\mathbf{B}$ is the magnetic field and $e_\textnormal{tot} = e_\textnormal{gas} + \frac{1}{2} \rho \mathbf{U}^2 + \frac{1}{2}\mathbf{B}^2$ is the total energy. The gas pressure is related to its internal energy via $P_\textnormal{gas} = (1 - \gamma)e_\textnormal{gas}$, where $\gamma$ is the adiabatic index. To achieve a de-facto isothermal equation of state, we set $\gamma = 1.0001$. We use a formally second-order accurate method consisting of Van-Leer type predictor-corrector integration scheme, the Harten-Lax-van Leer-Discontinuities (HLLD) Riemann solver \citep{HLLD}, as well as a piecewise-parabolic reconstruction method and mixed hyperbolic-parabolic divergence cleaning \citep{dedner2002}.
Note that in the absence of explicit dissipative terms, our simulations rely on numerical dissipation.
Thus, they are implicit large eddy simulations whose energy transfers match direct numerical simulations \citep{Grete2023}. 

\subsection{Initial conditions}
We added an interface to the \texttt{heFFTe} library \citep{Ayala2020} in \texttt{Parthenon} to initialize the magnetic field as a Gaussian random field with a double power-law spectrum:
\begin{equation}
    E_M(k) \propto k^{n_1} \cdot \left( 1 + \left(\frac{k}{k_I}\right)^\alpha \right)^{-(n_1 + n_2) / \alpha}
\end{equation}
such that $E_M(k)\propto k^{n_1}$ for $k < k_I$ and $E_k\propto k^{-n_2}$ for $k > k_I$. $\alpha = 10$ controls the sharpness of the transition. Following \citet{ refId0}, we focus on a Batchelor spectrum, i.e., $n_1=4$ with a sub inertial range of $n_2=2$. The initial peak of the spectrum is $k_{I,0}= 30$. Together with a box size of $2048^3$, this value gives us a large enough inertial range, but also sufficient scale separation between the integral scale and the box size that the inverse transfer has space to develop. 

Following \citet{Reppin2017NonhelicalTA}, we set the initial root-mean-square magnetic field strength to $0.3$ and the initial fluid pressure to $1$, both in internal code units. These values place the simulations in the nearly incompressible regime, where density fluctuations remain at the level of $\sim 1\%$. For the subsequent analysis, we therefore assume a solenoidal velocity field, i.e. $\nabla \cdot \mathbf{u} = 0$. 

The simulations are performed in a periodic box with side length $2\pi$. To ensure solenoidality and control the helicity, we set up the magnetic field from the vector potential in the helical mode decomposition and then take the curl to obtain the magnetic field. The velocity field is initially set to zero. We compare non-helical and helical simulations, i.e. $h=0$ and $h=1$ as defined in equation \ref{eq:frac_helicity}. Important parameters are summarized in table \ref{tab:sim_params}. 

\begin{table*}
\centering
\begin{tabular}{lc}
\hline 
Grid resolution & $2048^3$\\
Box size & $(2\pi)^3$ \\
Initial velocity field & $0$\\
Initial RMS magnetic field $B_{\rm rms}$ & $0.3$ \\
Initial fluid pressure $p$ & $1$\\
Magnetic power spectrum & Batchelor ($k^4$ subinertial range)\\
Fractional helicity & $h(k) = 0$ (non-helical) and $h(k) = 1 $ (helical) $\: \forall k$\\ 
\hline 
\end{tabular}
\caption{Summery of the simulation setup.}
\label{tab:sim_params}
\end{table*}

\section{Shell-to-shell transfer formalism}\label{sec:shell-to-shell}
To calculate the transfers, we use an open source energy transfer analysis framework \footnote{The framework implements the formalism described in \citet{10.1063/1.4990613}. It is available at \protect{\url{https://github.com/pgrete/energy-transfer-analysis}} and makes use of the \texttt{mpi4py-fft} library \citep{mpi4py-fft}.}.  
In the incompressible limit, i.e., with homogeneous pressure and density $\rho = 1$, the shell-to-shell energy transfer functions are \citep{PhysRevE.72.046301, DAR2001207, 2011PhPl...18k2307T},
\begin{equation}
\begin{aligned}\label{eq:energy_transfer}
& T_{\mathrm{BB}}(Q,K) = -\int \mathbf{B}_K (\mathbf{U}\cdot \nabla)\mathbf{B}_Q\, dV \\
& T_{\mathrm{UU}}(Q,K) = -\int \mathbf{U}_K (\mathbf{U}\cdot \nabla)\mathbf{U}_Q\, dV \\
& T_{\mathrm{UBT}}(Q,K) = \:\:\int \mathbf{B}_K (\mathbf{B}\cdot \nabla)\mathbf{U}_Q\, dV \\
& T_{\mathrm{BUT}}(Q,K) = \:\:\int \mathbf{U}_K (\mathbf{B}\cdot \nabla)\mathbf{B}_Q\, dV.
\end{aligned}
\end{equation}
These are derived from the MHD equations by decomposing the Fourier space into shell-bins $K_i = \{\mathbf{k}: k_i < |\mathbf{k}| <k_{i+1}\}$ and then defining the shell-decomposed real-space fields $\mathbf{F}_{K_i}(\mathbf{x})$ as the inverse Fourier transform of 
\begin{equation}
\mathbf{F}_{K_i}(\mathbf{k}) =
\begin{cases}
\mathbf{F}(\mathbf{k}), & \text{if } \mathbf{k}\in K_i \\
0, & \text{otherwise}
\end{cases}
\end{equation}
The type of binning that is chosen is important as it affects the structure of the transfer functions and the subsequent interpretation. Thin linear binning filters out spatially extended wave-like structures in real-space while logarithmic binning filters out structures that are localized in spectral and real space, i.e. eddies of different sizes \cite{Aluie2009}. Since we are interested in probing the interactions of spatially localized patches of helicity, we choose logarithmic (quarter-octave) binning with 33 shells in total. With binning understood, we drop the indices in subsequent notation. 

The transfer functions can be interpreted as follows: $T_{\textnormal{UU}}(Q,K)$ is the energy-transfer rate from velocity-shell $Q$ to velocity-shell $K$. The mechanism behind this transport is advection and is therefore mediated by the velocity-field. Similarly, $T_{\textnormal{BB}}(Q,K)$ is the transfer of magnetic energy from shell $Q$ to $K$ via advection. $T_{\textnormal{UBT}}(Q,K)$ is the energy-transfer rate from velocity-shell $Q$ to magnetic shell $K$ via the magnetic tension force, thereby mediated by the magnetic field. $T_{\textnormal{BUT}}(Q,K)$ is the inverse direction of this process. By definition, intra-reservoir transfer functions are antisymmetric, $T(Q,K) = -T(K,Q)$, while cross-reservoir transfer functions do not have this constraint. In the incompressible limit, this set of transfer functions is complete, i.e. the energy-change in each shell $K$ can be written as

\begin{equation}\label{eq:transfer_function_completeness}
\begin{aligned}
&\frac{d E_M(K)}{dt} = \sum_Q \: T_\mathrm{BB}(Q,K) + T_\mathrm{UBT}(Q,K) + \xi_M(K) \\
& \frac{d E_K(K)}{dt} = \sum_Q \: T_\mathrm{UU}(Q,K) + T_\mathrm{BUT}(Q,K) + \xi_K(K),
\end{aligned}
\end{equation}

where $\xi$ indicates energy-loss due to dissipation. In a similar fashion, we can define a transfer function for magnetic helicity\citep{Alexakis_2006, Plunian_Stepanov_Verma_2019, Teissier_Müller_2021}:
\begin{equation}\label{eq:helicity_transfer}
T_H(Q,K) = 2\int \mathbf{B}_K \left( \mathbf{U} \times \mathbf{B}_Q \right) \:dV
\end{equation}
Since there is no cross-reservoir helicity-transport, this transfer function is complete on its own, 
\begin{equation}
    \frac{dH(K)}{dt}= \sum_Q T_H(Q,K) + \xi_H(K),
\end{equation}
where $H(K)$ is the total helicity in shell $K$ and $\xi_H$ is the dissipation rate of magnetic helicity. 

\section{Results}\label{sec:results}

\begin{figure*}
    \centering
    \includegraphics[width=\linewidth]{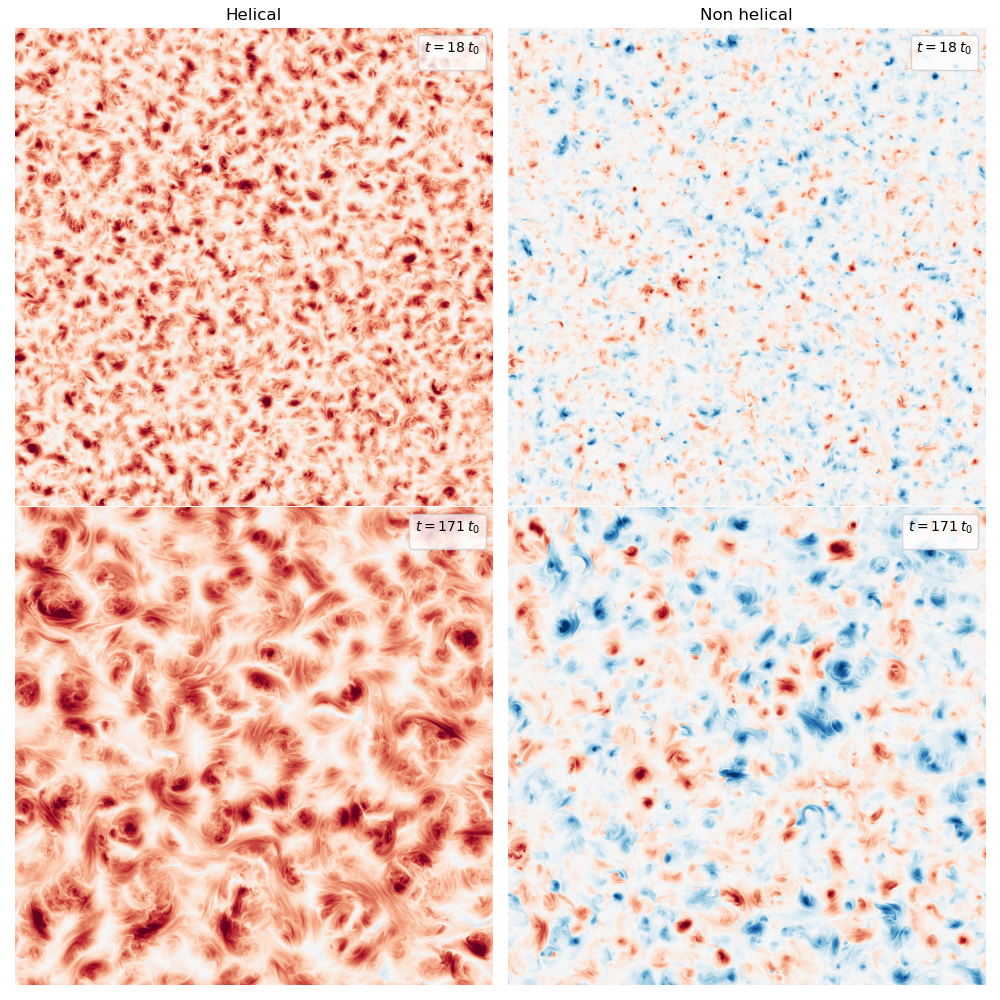}
    \caption{Slices of the helicity-density $\mathcal{H}(\mathbf{x})$ (calculated in Coulomb gauge) are shown for the helical (left column) and non-helical (right column) simulations at $t = 18\:t_0$ (upper row) and $t = 171\:t_0$ (lower row). Colors indicate the value of $\mathcal{H}(\mathbf{x})$ (blue = negative, red = positive) and a linear color-scale is used. For each subplot, the field is normalized to its maximum individually.}
    \label{fig:Helicity_density}
\end{figure*}

In this section, we present the results from our simulations. Their setup is summarized in table \ref{tab:sim_params}. Throughout the analysis, we use the Coulomb gauge to recover the vector potential from the magnetic field and calculate magnetic helicity. Inspired by the dynamical picture presented in HS, we plot slices of the helicity-density $\mathcal{H}(\mathbf{x})$ for the helical and non-helical simulations at earlier and later times in Fig. \ref{fig:Helicity_density}. As expected, the helical field only consists of positively helical patches, while for the non-helical field, positively and negatively helical patches are equally abundant. 
They are localized in space and their typical correlation length grows with time due to inverse transfer. 
In the supplemental material, we include an animated version of this plot, which has been created using a box size of $1024^3$ cells and an initial integral scale of $k_I = 20$.

In the rest of this section, we first analyze the spectral evolution of magnetic and kinetic energy and magnetic helicity, followed by a discussion of the corresponding energy‑transfer functions. From these transfers we measure characteristic timescales for inverse energy transfer mediated by magnetic–magnetic (BB) and kinetic–magnetic (UBT) interactions. Finally, we assess the couplings of different scales in the positively and negatively helical parts of the field by evaluating energy and helicity transfers in the helical‑mode decomposition.

Throughout the paper, time is expressed in units of initial Alfven times, $t_0 = (v_{A,0}\cdot k_{I,0})^{-1}$, where $v_{A,0}$ is the initial Alfven velocity and $k_{I,0}$ is the initial integral (energy-containing) wavenumber. We define $k_I$ as the wavenumber where $E_M(k_I)$ is largest. The integral scale is the inverse of this, $L = 1/k_I$.

\subsection{Spectra}\label{subsec:spectra}
\begin{figure*}
    \centering
    \includegraphics[width=\linewidth]{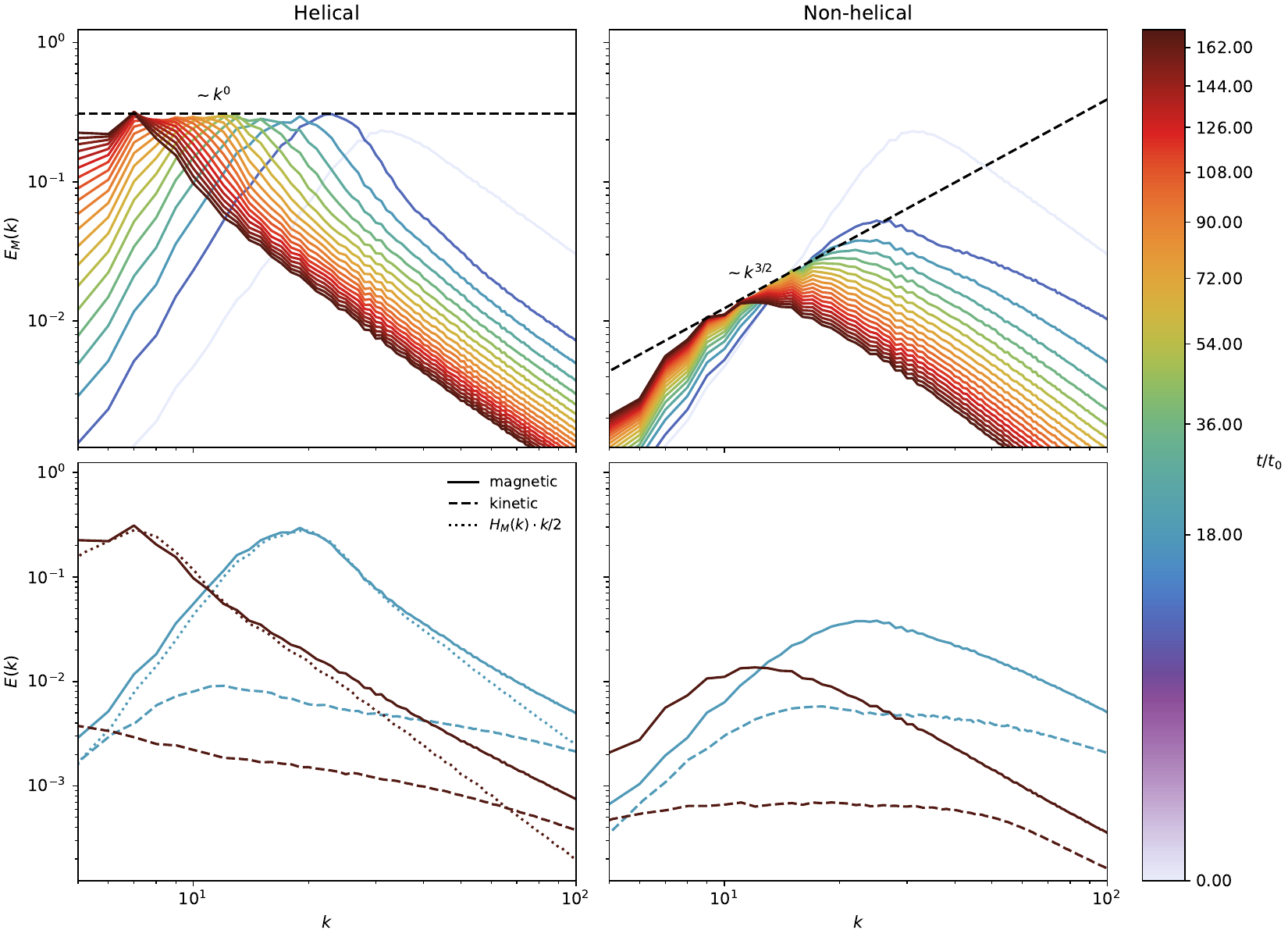}
    \caption{Top row: The evolution of the magnetic energy spectra is shown for the fully helical (left panel) and non-helical (right panel) runs. Time is given in units of initial Alfven times $t_0 = (v_{A,0} \cdot k_{I,0})^{-1}$, where $v_{A,0}$ is the initial Alfven velocity and $k_{I,0}$ is the initial integral wave-number. Dashed lines indicate the expected evolution of the peak for the conservation of magnetic helicity in the helical case and the Hosking integral in the non-helical case.
    Bottom row: The magnetic (solid) and kinetic (dashed) power spectra are shown for early ($t/t_0 = 18$) and late ($t/t_0 = 171$) times. For helical (left panel) and non-helical (right panel) initial conditions. Additionally, for the fully-helical case, the quantity $\mathcal{H}(k) \cdot k/2$ is plotted (dotted line), which, according to equation \ref{eq:realizability} indicates the degree of helicity-saturation.}
    \label{fig:spectra}
\end{figure*}

In the upper panels of \ref{fig:spectra}, we show the evolution of the magnetic energy spectra for the helical and non-helical runs. Inverse energy transfer is clearly evident in both cases since the peak is moving to the left. However, as expected, it is much more pronounced in the fully helical case. Crucially, both spectra evolve approximately self-similarly in the sense that their shape stays approximately the same. In particular, the growth of the $k< k_I$ modes, which we will call the infrared (IR) range, is multiplicative, i.e. 

\begin{equation}\label{eq:multiplicative_growth}
\frac{1}{E_M(k)}\frac{d E_M(k)}{dt} :=\frac{1}{\tau(t)} ,
\end{equation}
with a timescale, $\tau(t)$, that does not depend on $k$. 
The evolution of the spectral peak as a function of the integral scale is set by the relevant conserved quantity, which follows from dimensional arguments: For helicity ($H\sim E_ML$), one expects $E_{M, \text{peak}}(t)\sim E_M(t) L(t) \propto k_I(t)^0$, while the Hosking integral ($I_H\sim E_M^2 L^5$) implies $E_{M, \text{peak}}(t)\propto k_I(t)^{3/2}$ \cite{atmos14060932}. We see that our simulations follow these expected trends after an initial "settling phase" of $t \approx 18\: t_0$. To ensure that we are probing the fully developed turbulent regime, we calculate our energy and helicity transfer functions at this time. They are discussed in the following section. 

In the lower panels of figure \ref{fig:spectra}, we also compare the magnetic energy spectra with the kinetic ones. We see that the system is magnetically dominated at all times. Additionally, for the helical case, we plot the quantity $\mathcal{H}(k) \cdot k/2$, which, according to equation \ref{eq:realizability} indicates the degree of helicity-saturation. We find that the system, even though it started out fully saturated, looses some saturation at small and large scales, but remains saturated close to the integral scale. 

\subsection{Energy transfer}

Having established that the inverse transfer scales according to the conservation in the maximally helical case and the Hosking integral in the non-helical case, we now investigate how this transfer is realized in terms of shell-to-shell interactions.
To this end, the energy-transfer functions (equation \ref{eq:energy_transfer}) are computed from the simulations at $t = 18 \: t_0$.

\begin{figure*}
    \centering
    \includegraphics[width=0.8\linewidth]{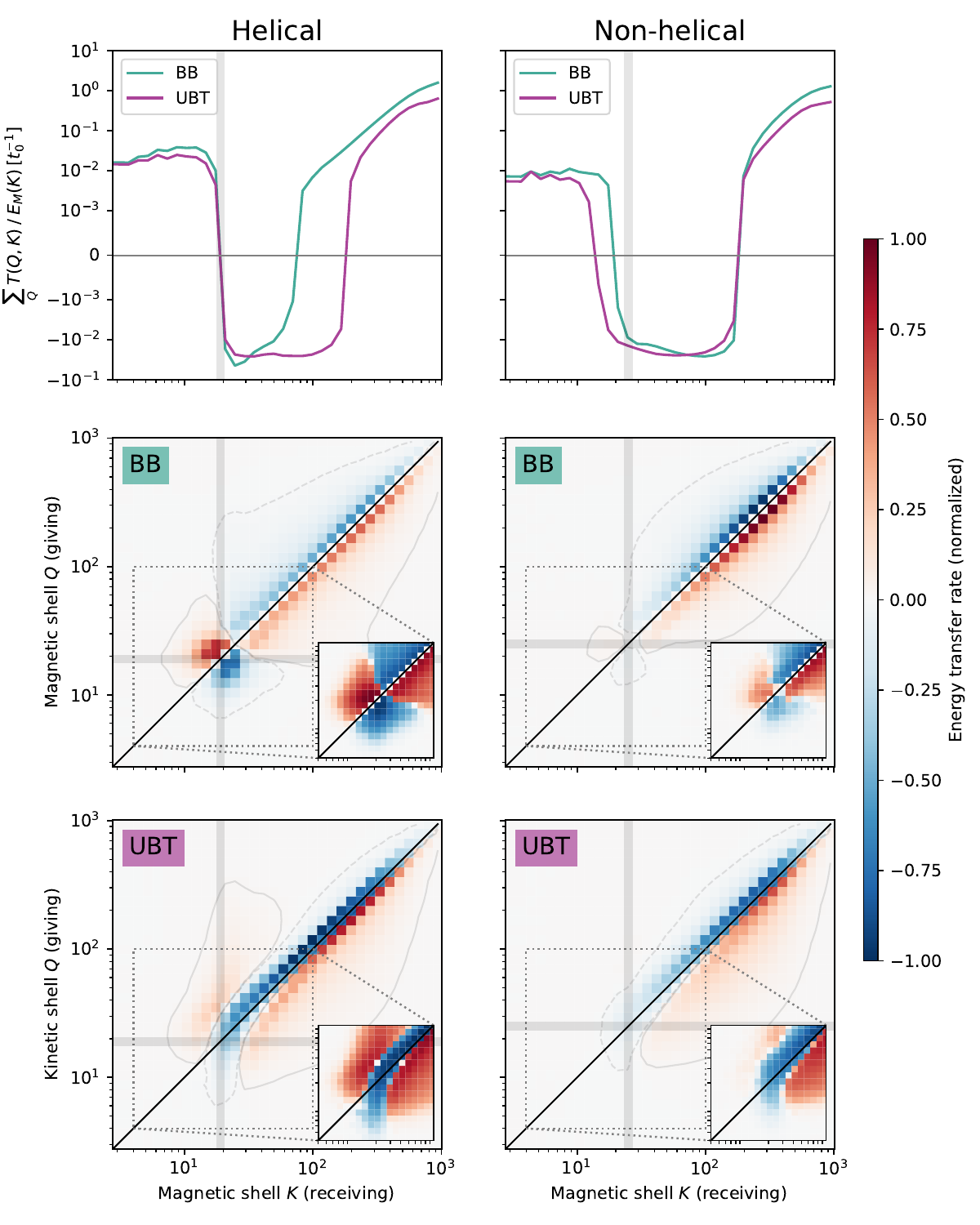}
    \caption{The energy transfer terms for transfer into the magnetic reservoir ($T_{\mathrm{BB}}$ and $T_{\mathrm{UBT}}$ in Eq. \ref{eq:energy_transfer}) are plotted as heatmaps for the helical (left column) and non-helical (right column) cases at $t = 18 t_0$. The bottom row shows heatmaps for the energy-transfer from the velocity reservoir into the magnetic reservoir via magnetic tension ($T_{UBT}$) and the middle row shows the energy transfer within the magnetic reservoir via advection ($T_{BB}$). Receiving shells are plotted on the horizontal axis and donating ones on the vertical axis. The heatmaps are normalized by the maximum over both the BB and UBT channels for each column individually. The main heatmaps are plotted in a linear color-scale, as indicated by the colorbar. The insets show the region around the integral scale using a symmetric logarithmic color-scale with a linear threshold at $1\%$ of the maximum value. The contour-lines trace the regions where the transfer rate is $1\%$ of the maximum. The horizontal/vertical gray lines indicate the integral scale.
    For each transfer channel $X\in \{\mathrm{BB}, \mathrm{UBT}\}$, the resulting instantaneous growth rates of each shell $\tau_X^{-1}(K) = \sum_Q T_X(Q,K)/E_M(K)$ are plotted in the top row.}
    \label{fig:transfer_maps}
\end{figure*}

In Fig. \ref{fig:transfer_maps}, we plot the transfer functions that correspond to transfer into the magnetic reservoir, i.e. $T_\text{BB}(Q,K)$ and $T_\text{UBT}(Q,K)$. \footnote{Since $T_\textnormal{BB}(Q,K)$ is asymmetric in $Q$ and $K$, the values at the diagonal are expected to be zero. Non-zero values across the diagonal are thus due to numerical errors, which are primarily caused by taking the gradients in equation \ref{eq:energy_transfer} in real space on a discrete mesh. They therefore only meaningfully affect the high-$k$ modes.}.

We will start by discussing the qualitative features in the transfer maps, which are similar between helical and non-helical cases, and, due to the self-similar evolution of the system, also independent of time after turbulence has fully developed. 

For shells above the integral-wave-number, $T_{\text{BB}}(Q,K)$ shows a direct cascade: Energy is transferred locally from the integral scale towards smaller scales. Constant values close to the diagonal for $k>k_I$ indicate a constant energy-flux across the inertial range, as is typical for a Kolmogorov-type direct cascade.

Positive values above the diagonal in $T_\mathrm{BB}(Q,K)$ indicate inverse energy transfer, which is strongest for $Q,K\sim k_I$ and gradually gets weaker for smaller $K$ and larger $Q$. 

The $T_\mathrm{UBT}$ transfer-functions show that at $K,Q > k_I$, the magnetic field looses energy to kinetic shells identical (negative transfer on the diagonal) or slightly smaller than itself but gains energy from larger kinetic shells. The inertial-range kinetic and magnetic modes thus undergo a \textit{coupled direct cascade}, in which energy is transferred from large to small scales not just within one reservoir, but also across reservoirs \citep{Grete2021tension}.

From the inset in the helical case and the resulting rates in the top row of the figure, it is evident that kinetic-magnetic inverse transfer is also present which is similar (albeit slightly weaker) in magnitude to the magnetic-magnetic transfer. 

Negative values of $T_\mathrm{UBT}(Q,K)$ for $Q, K\sim k_I$ imply magnetic-to-kinetic energy transfer close to the integral scale. 

The negative region below the diagonal in the UBT map implies energy transfer from the integral-scale magnetic field to large-scale velocity shells. This indicates that the growth of magnetic and velocity-structures is tightly coupled, which is expected for a magnetically dominated system. 

\begin{figure}
    \centering
        \includegraphics[width=\linewidth]{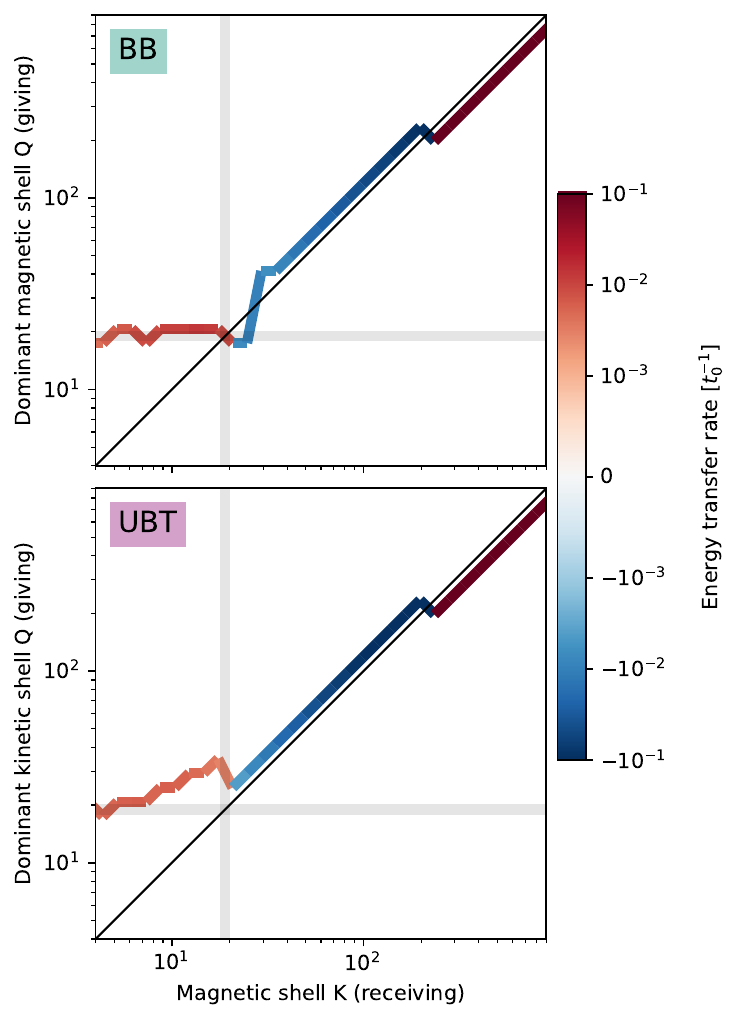}
    \caption{For each receiving magnetic shell $K$ in the helical simulation, the dominant donating shell $Q$ is plotted for the $T_{\text{BB}}$ (upper panel) and $T_{\text{UBT}}$ (lower panel panel) transfer functions. The color indicates the corresponding transfer rate, $T(Q_\mathrm{dom}, K)/E_K$, where $E_K$ is the energy in the receiving shell, in units of initial Alfven times, $t_0$. 
    The diagonal line indicates transfer between identical scales. The light-gray vertical and horizontal lines indicate the integral scale.}
    \label{fig:Q_shells}
\end{figure}

To make it more apparent which scale-interactions dominate, we plot the dominant donating shells for each receiving shell in the magnetic reservoir for the Helical simulation in figure \ref{fig:Q_shells}. It is apparent that, regardless of the channel, all IR-shells get their energy directly from the integral scale. Consequently, inverse transfer is increasingly non-local for IR-shells with lower $k$. 

\begin{figure}
    \centering
    \includegraphics[width=0.7\linewidth]{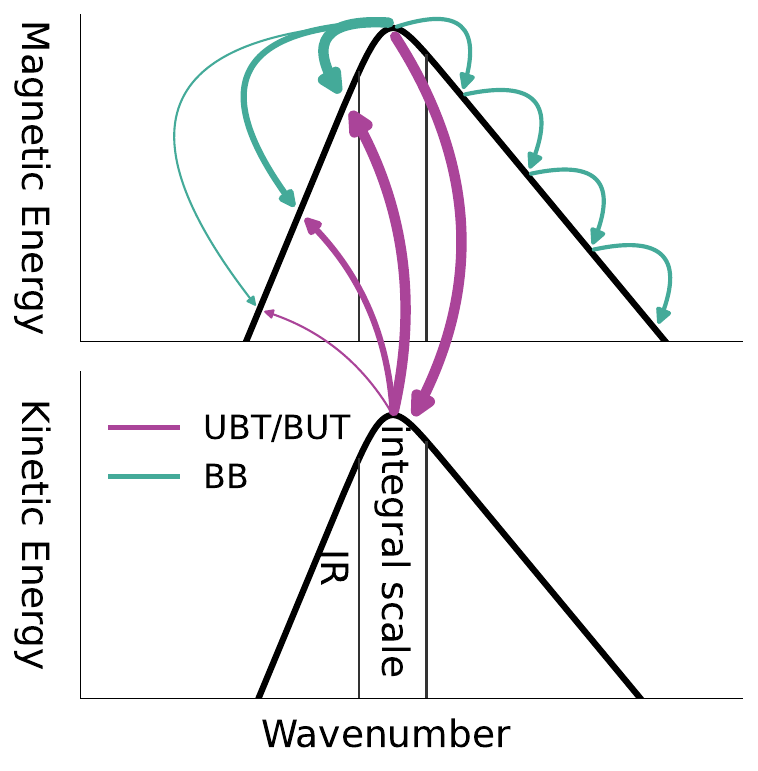}
    \caption{Schematic of the energy flow for both helical and non-helical cases: Arrow thickness indicates relative transfer rate (not to scale).
    Within the magnetic reservoir (teal arrows), there is both local direct transfer, as indicated by arrows going from the integral scale to the right, as well as local and non-local inverse transfer, as indicated by arrows going to the left. Inverse transfer is always coming from the integral scale and thus gets progressively more non-local, and also weaker, for larger receiving scales. Purple arrows indicate cross-reservoir transfer (UBT/BUT channel). Energy flows from the magnetic to the kinetic integral-scale and is then, as in the BB channel, distributed into the magnetic IR at a rate that is proportional to the energy in the magnetic IR-shells. For simplicity, transfer that leads to changes in the kinetic reservoir is not shown.}
    \label{fig:energy_transfer_schematic}
\end{figure}

We thus identify two main channels for energy transfer into the magnetic IR, as sketched in figure \ref{fig:energy_transfer_schematic}. Inverse transfer via the BB-channel corresponds to energy being transported from the magnetic integral scale directly to magnetic IR-modes. In the UBT channel, transfer proceeds from the magnetic integral scale to the kinetic integral scale and then to the magnetic IR shells. For both channels, inverse transfer leads to a constant growth rate $E_M(K)\:( d E_M(K)/dt)^{-1}$ for $K< k_I$, as evident from the top panels in Fig. \ref{fig:transfer_maps}. This multiplicative growth is a direct consequence of the self-similar spectral shift seen in Fig. \ref{fig:spectra}.  

The underlying triadic Fourier-mode interactions giving rise to energy transfer into the IR are therefore given by $\mathbf{k}_{IR} = \mathbf{p} + \mathbf{q}_{I}$, where $\mathbf{p}$ is the mediating mode. Since $|\mathbf{k}_{IR}|\ll |\mathbf{q}_I|$, this implies $|\mathbf{p}|\approx |\mathbf{q}_I|$, i.e. both the donating and mediating modes responsible for energy-transfer into the IR are integral-scale sized. We have also tested this directly by decomposing the mediating fields in $T_\mathrm{BB}(Q,K)$ and $T_\mathrm{UBT}(Q,K)$. 

\subsection{Total energy transfer timescales}\label{sec:timescales}

\begin{figure}
    \centering
    \includegraphics[width=1\linewidth]{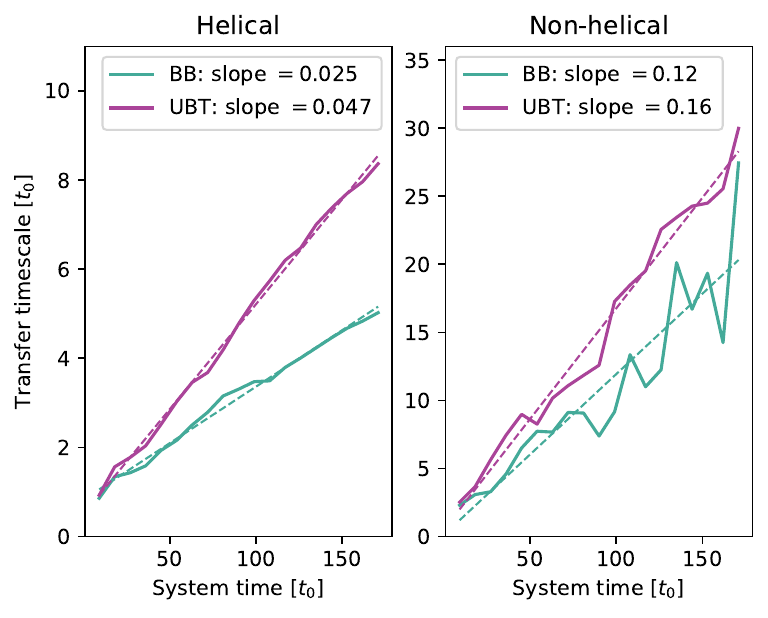}
    \caption{The inverse-transfer timescales as defined in equation \ref{eq:channel_growth_rates} are averaged over IR-shells and plotted as a function of system time.
    The dotted lines show a linear fit to the data with the slope given in the legend.
    }
    \label{fig:inverse_transfer_timescales}
\end{figure}

Since the instantaneous growth rates 

\begin{equation}\label{eq:channel_growth_rates}
\tau_X^{-1}(K) = \sum_Q T_X(Q,K)/E_M(K)\:\mathrm{for} \:X\in\{\mathrm{BB}, \mathrm{UBT}\}
\end{equation}
are constant across the magnetic IR, as shown in the top row of Fig. \ref{fig:transfer_maps}, they define a scale-independent inverse-transfer timescale $\tau_X(t)$ for each channel. Since BB and UBT form a complete set of transfer functions, i.e. equation \ref{eq:transfer_function_completeness} is satisfied and dissipation is negligible for IR-shells, the total inverse-transfer timescale is given by $\tau_I = (\tau_{\mathrm{BB}}^{-1} + \tau_{\mathrm{UBT}}^{-1})^{-1}$.
In practice, we average $\tau_{\mathrm{BB}}(K)$ and $\tau_{\mathrm{UBT}}(K)$ over the IR by taking the mean between  shells with $k\in (3.36, 5.66)$. In Figure \ref{fig:inverse_transfer_timescales}, we plot the timescales obtained in this way for the helical and non-helical cases as a function of simulation-time $t$. We find a simple linear relation, which is expected if each IR mode grows as a power-law with time. By measuring the slope of $\tau(t)$, we obtain the power-law index for the inverse-transfer. Comparing the relative contributions of the different channels, we find $\tau_{\mathrm{BB}} / \tau_{\mathrm{UBT}} \approx 0.52$ in the helical case and $\tau_{\mathrm{BB}} / \tau_{\mathrm{UBT}} \approx 0.72$ in the non-helical case. The relative contribution of the UBT-channel is thus smaller in the helical case, but still significant. Furthermore, we compare the total inverse-transfer timescales in the helical and non-helical case and find $\tau_\textnormal{helical}/\tau_\textnormal{non-helical} \approx 0.26$. In Sec. \ref{subsec:helicity_decomp}, this ratio is interpreted in terms of different helical mode couplings.

We can also express these timescales relative to the Alfven time $\tau_A(t) = [v_A(t)\cdot k_I(t)]^{-1}$, which also scales linearly as a function of system time. We obtain an average of $\tau_I/\tau_A\approx 4.53$ for the helical case, and $14.29$ for the non-helical one. 

\subsection{Helicity transfer in the fully helical case}
\begin{figure}
    \centering
    \includegraphics[width=\linewidth]{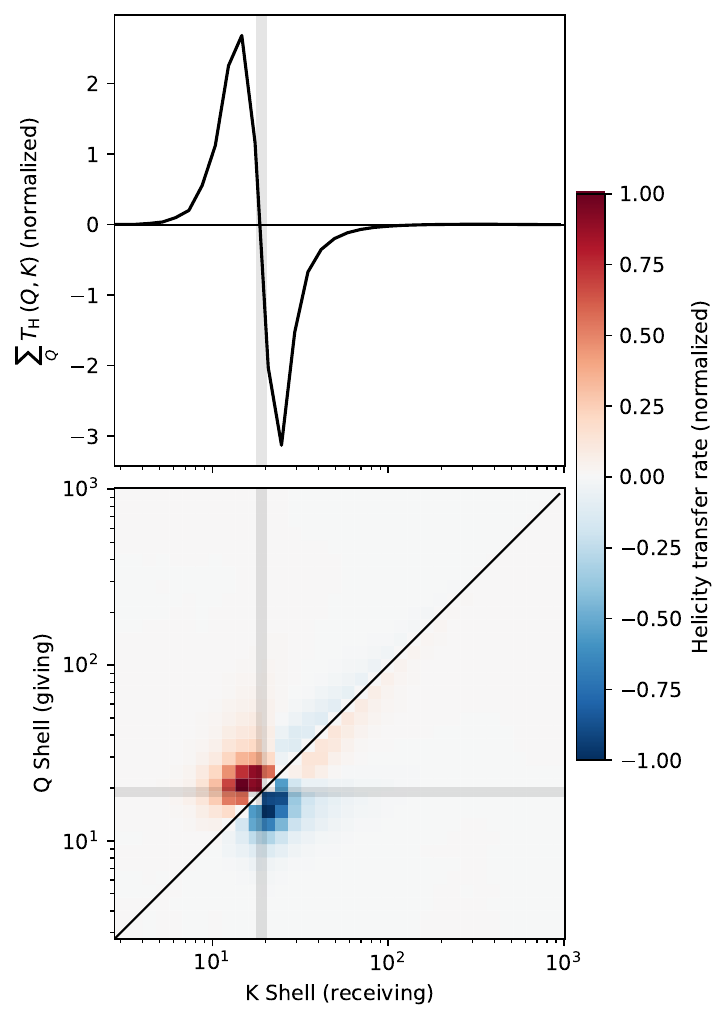}
    \caption{The helicity-transfer function as defined in Equation \ref{eq:helicity_transfer} is shown for the fully-helical run at $t = 18\tau_0$, normalized to its maximum. The upper panel shows the total helicity-change in each receiving mode $k$, i.e. the sum over all donating modes.}
    \label{fig:hel_transfer}
\end{figure}

Since helicity plays a major role in inverse energy transfer, it is a natural next step to investigate the shell-to-shell transfer function for magnetic helicity (Eq. \ref{eq:helicity_transfer}) and its relation to energy transfer. In Fig. \ref{fig:hel_transfer}, we plot it for the fully-helical case. We find strong inverse transfer, qualitatively similar to the BB transfer function, but almost no direct transfer. 
This is expected from the realizability condition $|\mathcal{H}(k)|\leq 2 E_M(k)/k$, introduced in Sec. \ref{sec:definitions}, which states that the amount of helicity per unit energy that can be stored at a scale $k$ decreases with larger $k$. If helicity is already (close to) saturated at large $k$, as it is in the fully-helical case, then direct helicity transfer is strongly constrained. \footnote{Note that even for the initially fully helical setup, helicity is no longer fully saturated at large $k$ at later times, as can be seen in Fig. \ref{fig:spectra}, explaining why a very small amount of direct helicity-transfer is still possible.}

\subsection{Helicity-decomposed transfer functions in the non-helical case}\label{subsec:helicity_decomp}

\begin{figure*}
    \centering
    \includegraphics[width=0.75\linewidth]{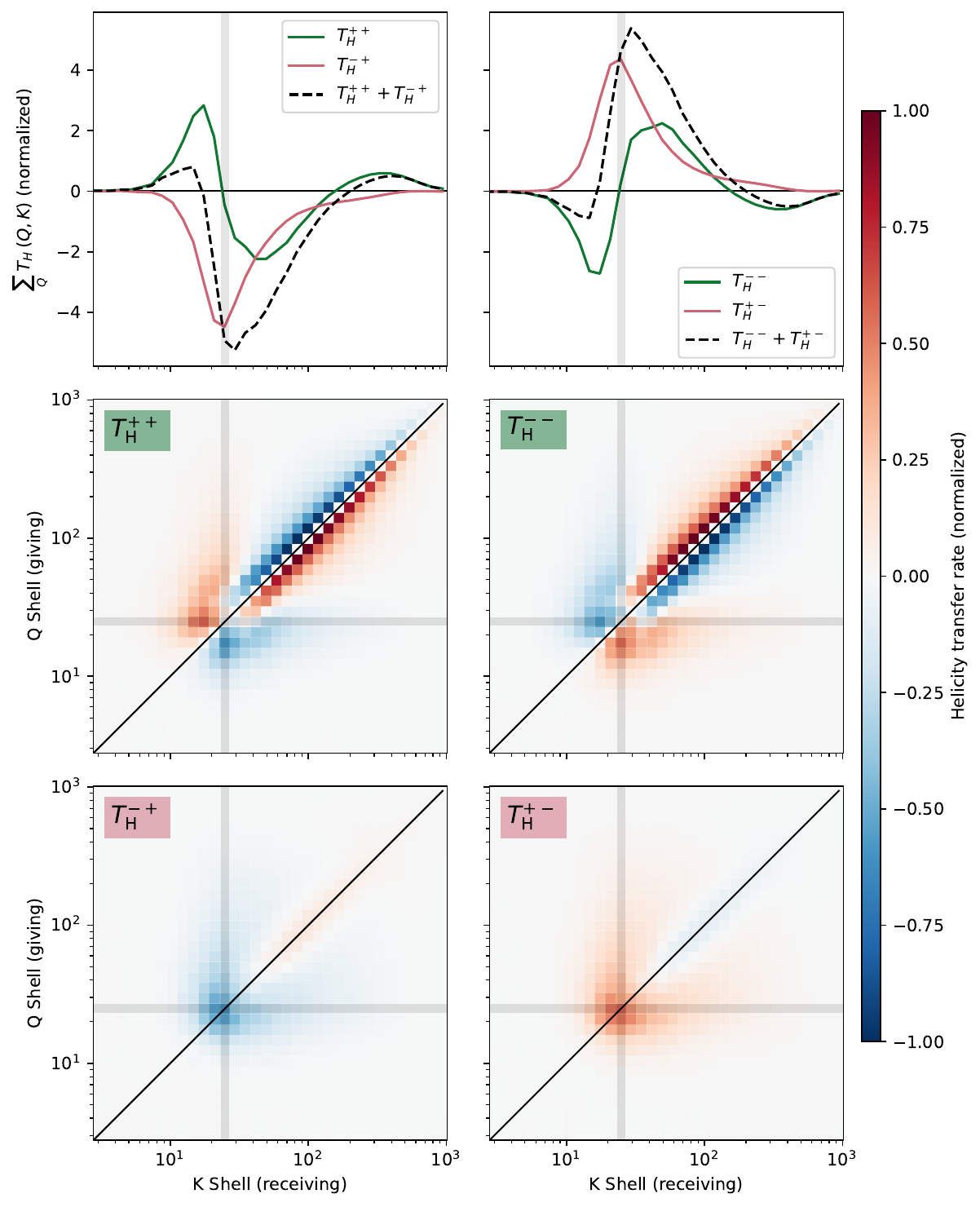}
    \caption{The decomposed helicity-transfer functions as defined in equation \ref{eq:decomp_hel_transfer} are shown for the non-helical run at $t=18\:t_0$. Transfer functions corresponding to changes in the positively and negatively helical part of the field are shown in the left and right columns respectively. The bottom and middle rows show contributions from mixed and equal interactions respectively. The top row shows the corresponding net changes, i.e. the sum over all donating shells. Transfer function are normalized their maximum value across all helicity- and shell-combinations.}
    \label{fig:decomp_hel_transfer}
\end{figure*}

In the non-helical case, the energy of the magnetic field is distributed evenly between the positively and negatively helical parts of the field, as defined in equation \ref{eq:helical_decomp}. Since we expect transfer of positive and negative helicity to cancel out on average, calculating the helicity transfer function for the non-helical field in the same way as we did for the fully helical one would be meaningless. We therefore analyze them in the helical mode decomposition, isolating inverse transfer in the positively and negatively helical parts of the field. For the transfer function of magnetic helicity (equation \ref{eq:helicity_transfer}) we get four separate contributions,

\begin{equation}\label{eq:decomp_hel_transfer}
T_H^{\sigma_1\sigma_2}(q,k) = 2 \int \mathbf{B}_p^{\sigma_2} \cdot 
  \left( \mathbf{U} \times \mathbf{B}_q^{\sigma_1} \right) \, dV, 
\end{equation}
where $\sigma_1, \sigma_2\in \{+,-\}$ are the signs of the donating and receiving parts of the field respectively. We plot these for the non-helical run in Fig. \ref{fig:decomp_hel_transfer}. 

$T_H^{++}$ measures the helicity-transfer between shells in the positively-helical part of the field.
Contrary to the fully-helical case, this transfer map shows a pronounced local direct helicity cascade for shells above the integral scale in addition to the inverse transfer. Since the helicity-content in both helical sectors averages to zero at each scale, $\mathcal{H}(k)\approx 0$, the realizability condition (Eq. \ref{eq:realizability}) does not constrain the direct transfer in each sector. 

We furthermore find $T_H^{--} = -T_H^{++}$, since in the case of vanishing total helicity, negatively and positively helical structures should be equally abundant, as can also be seen in figure \ref{fig:Helicity_density}, and their dynamics will be the same. Mergers of negatively helical structures lead to inverse transfer of negatively signed helicity.

$T_H^{+-}$ and $T_H{-+}$ correspond to cross-sector helicity transfer. $T_H^{-+}$ is negative around the integral scale in both $K$ and $Q$, while $T_H^{+-}$ is positive. They do not integrate to zero and thus correspond to a net loss/gain of helicity in the respective sectors, as can also be seen in the Q-integrated transfer functions in the upper panels of Fig. \ref{fig:decomp_hel_transfer}. 

We interpret this as signed helicity being lost due to mutual annihilation of structures with opposite-signed helicity, as explained below. 

Assume a merger of two structures with equal and opposite helicity, resulting in a structure of zero helicity, which, according to HS is assumed to decay quickly. To understand the transfer functions, one needs to track the changes of helicity in the positively helical ($\mathbf{B}_+$) and negatively helical ($\mathbf{B}_-$) sectors separately. After the merger, $\mathbf{B}_+$ has lost helicity, while $\mathbf{B}_-$ has gained helicity, since it is now less negative. Thus, $T_H^{+-}$ will be positive, while $T_H^{-+}$ will be negative.

Annihilation events therefore decrease the absolute helicity in each sector. According to equation \ref{eq:signed_helicity_energy_relation}, the absolute helicity and magnetic energy in each sector are directly related. Consequently, loss of absolute helicity in a given sector implies loss of magnetic energy. The consequence of this is that annihilation acts as a sink for magnetic energy, converting it into kinetic energy or heat.

Since not all merging structures will be of the same size, helicity is not only removed from the integral-scale but also from neighboring scales. The transfer functions from equal and opposite sign helicity mergers thus partially cancel, leading to less overall inverse transfer. This can be seen when comparing the green and dashed black lines in the upper panels of figure \ref{fig:decomp_hel_transfer}. 

To assess how these interactions affect the final energy-transfer, we apply the same decomposition to the energy transfer function for the magnetic reservoir, resulting in
\begin{equation}
\begin{aligned}\label{eq:energy_transfer_decomp}
& T_{\mathrm{BB}}^{\sigma_1 \sigma_2}(q,k) = -\int \mathbf{B}_k^{\sigma_2} (\mathbf{U}\cdot \nabla)\mathbf{B}_q^{\sigma_1}\, dV
\end{aligned}
\end{equation}
with $\sigma_1, \sigma_2 \in \{+,-\}$. We find $T_{\mathrm{BB}}^{++} = T_\mathrm{BB}^{--}$ and $T_\mathrm{BB}^{+-} = T_\mathrm{BB}^{-+}$, which is expected due to the symmetry between positive and negative helicity modes in the non-helical case. 

\begin{figure}[h!]
    \centering
    \includegraphics[width=\linewidth]{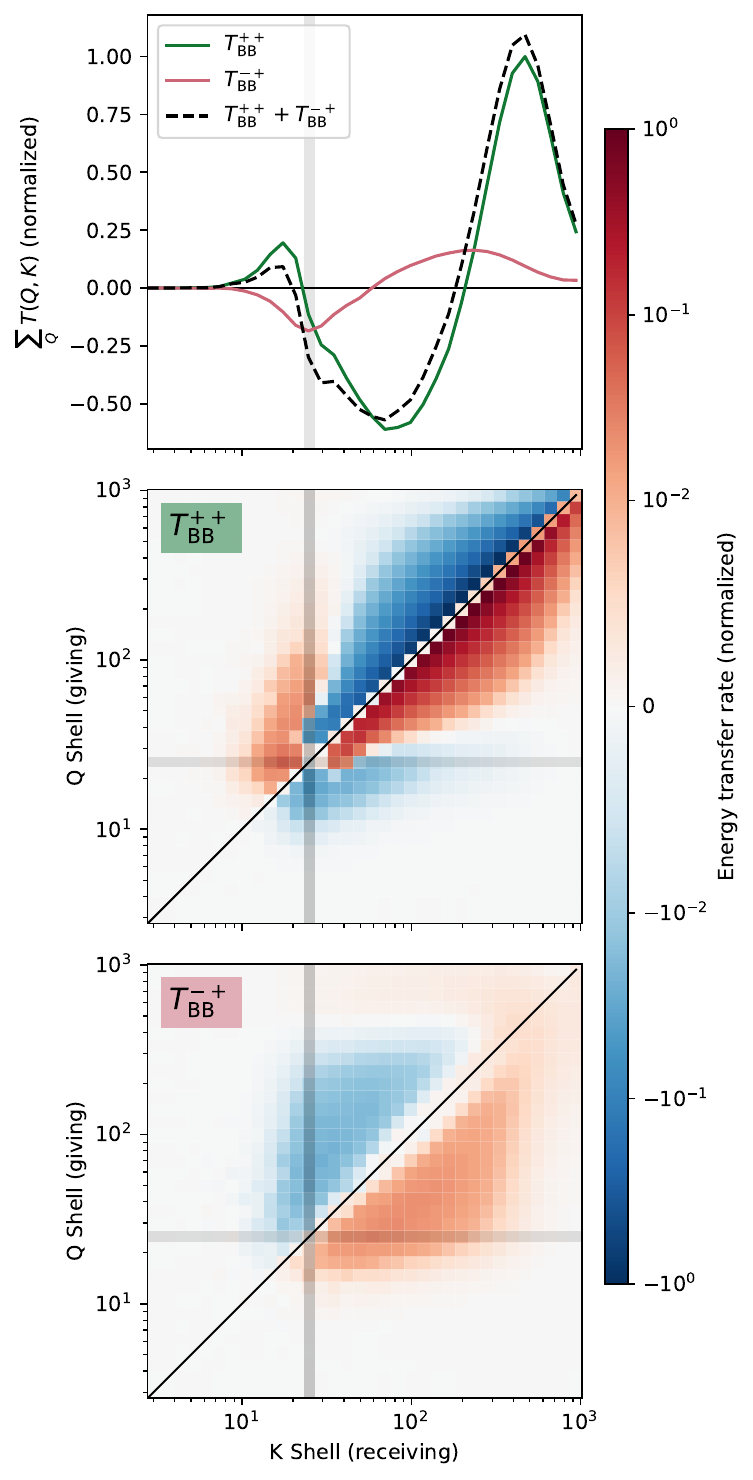}
    \caption{The decomposed energy-transfer function as defined in equation \ref{eq:energy_transfer_decomp} is shown for the non-helical case at $t = 18\:t_0$. Since it is invariant under a simultaneous sign change of $\sigma_1$ and $\sigma_2$, we only show it for $\sigma_1 = + $. The bottom panel shows mixed-helicity interactions, while the middle panel shows equal-helicity interactions. The top panel shows the resulting net changes in each receiving shell, i.e. the sum over all donating shells. The transfer function is normalized to its maximum across all helicity- and shell-combinations.}
    \label{fig:hel_decomp_energy_transfer}
\end{figure}

In Figure \ref{fig:hel_decomp_energy_transfer}, we therefore only show the combinations that lead to transfer into $\mathbf{B}_+$. $T_\mathrm{BB}^{++}$ is qualitatively similar to the total BB transfer functions, showing inverse transfer below and a direct cascade above the integral scale. However, the cross-sector transfer function $T_\mathrm{BB}^{+-}$ only shows direct transfer, which is much less scale-local than the direct cascade within each helical sector. This moves energy to small scales much more quickly, aligning with the picture of rapid decay of local magnetic energy after an annihilation event. 

As for helicity, energy lost due to opposite-sign interactions partially cancels the inverse transfer from equal-signed interactions, reducing the net inverse transfer in each sector, as indicated by the dashed black curve in the upper panel of Fig. \ref{fig:hel_decomp_energy_transfer}. 

We can use this fact to interpret the factor 0.26 difference between the inverse-transfer timescales in the helical and non-helical cases that we found in section \ref{sec:timescales}: In the simplified scenario where all islands have the same size (which would correspond to the case of a very narrow power-spectrum), we expect a factor 0.5 difference between the timescales, since same-sign and opposite-sign mergers are equally likely. If the power-spectrum is more spread out, more cancellation occurs, reducing this factor. This implies that non-helical inverse transfer will be less efficient for power spectra with a shallower IR-range, a fact which has been confirmed by simulations, e.g. \citet{Reppin2017NonhelicalTA}.

\section{Discussion}\label{sec:discussion}

In this section, we discuss the physical mechanism for inverse transfer that is implied by our diagnostics, and draw comparisons to previous works, particularly the theory presented in HS. This theory explains inverse transfer as the merging of spatially localized magnetic islands of positive or negative helicity-density. 

\textbf{Inverse transfer as island mergers.} For simplicity, we start by discussing the fully-helical case, i.e. we assume that all islands have positive helicity. The typical size of such an island is given by the integral scale, implying that the merger with the highest probability will be between two integral-scale sized islands. In the idealized scenario, such a merger would remove energy from the integral scale $L$ and transport it to a slightly larger scale. This explains why, at any given time, most inverse-transfer is roughly localized at the integral scale (both for the donating and the receiving scales). 

However, our diagnostics show that inverse transfer becomes increasingly non-local for structures larger than the integral scale: The dominant donating scale is always the integral-scale and energy is distributed non-locally to each IR-shell at a rate proportional to its energy, leading to self-similar multiplicative growth.   

Mergers would require magnetic reconnection to rearrange the local magnetic field topology. Reconnection usually results in a sudden transfer of energy from the magnetic into the kinetic reservoir, which, as has previously been noted by \citet{10.1093/mnras/staa3849}, would explain the scale-local magnetic-to-kinetic energy transfer that we observed. Outflows from reconnection sites may subsequently stretch the magnetic field lines to larger scales, performing work against magnetic tension in addition to advecting them. This would explain the observation that both magnetic-magnetic and kinetic-magnetic energy-exchange contributes to large-scale magnetic growth. Nevertheless, further parameter studies are needed to explain the relative contributions of these channels and how they vary with the parameters of the system, such as helicity content and diffusion parameters.

\textbf{Helical-mode decomposition.} 
In the case of vanishing net-helicity, we calculated energy and helicity transfer functions between and within the positively and negatively helical sectors. Close to the integral scale, equal- and opposite sign interactions are comparable in magnitude, aligning with the idea by HS that mergers of equal-and opposite helicity structures should have the same probability. While equal-sign interactions lead to inverse transfer of energy, opposite-sign interactions act as a sink for magnetic energy. \citet{Linkmann_Berera_McKay_Jäger_2016} carried out a purely analytical linear stability analysis of these mode-couplings. The idea behind this approach is that transfer functions which are linearly unstable should lead to energy transfer away from the receiving mode, while stable solutions should lead to energy transfer toward it. They concluded that interactions between like-helicity modes can cause non-local inverse transfer in magnetically dominated systems, aligning with our results. \citet{Teissier_Müller_2021} also measured decomposed helicity-transfer functions for a turbulence simulation with large-scale energy and small-scale helicity driving. They report inverse transfer of helicity due to magnetic stretching induced by solenoidal velocity modes at the integral scale. They also find that inverse helicity transfer is associated with magnetic-magnetic and kinetic-magnetic energy exchanges. 

\textbf{Dependence on IR-scaling.} In this work, we focused on a Batchelor magnetic energy spectrum, i.e. a spectrum with a $k^4$ sub inertial range. This scaling is the natural expectation for causally generated PMF, while steeper spectra are forbidden by causality constraints \cite{Ruth_Durrer_2003}. For shallower $k^2$ spectra, inverse transfer has not been found in the past \cite{Reppin2017NonhelicalTA}. From the Taylor expansion of the magnetic energy power spectrum (see HS), one expects that in this case, the Saffman integral $I_S = \int d^3r \left<B(x) B(x+r)\right>$ \cite{2023JPlPh..89f9006B} should replace the Hosking integral as the dynamically relevant conserved quantity, which would explain the lack of inverse transfer. However, \citep{2023JPlPh..89f9006B} report evidence of inverse transfer even in the $k^2$ case, albeit not conclusively due to resolution limitations.
Since shallower $k^3$ spectra might be realized by magnetogenesis during the electroweak phase transition \cite{Vachaspati_2021}, it would be interesting to redo diagnostics used in this paper for the shallower case as well. The conservation of the Hosking integral is tied to the dynamical picture of merging spatially localized helical patches. A shallower sub-inertial range would introduce more large-scale correlation, which might cause this picture to break down. 

\textbf{Timescales.} It should be noted that our diagnostics, while supporting the general dynamical picture presented in HS, they do not support their claim that energy should decay at the reconnection timescale. This is because the reconnection-timescale is derived from the simple sweet-Parker model of reconnection, which, as has been noted by \citet{ refId0}, might not directly translate to a turbulent system. In fact, during our convergence tests (see appendix \ref{sec:convergence}), we have found that increasing the resolution all else equal, thereby increasing the effective Lundquist number, does not change the decay timescale $\tau^{-1} = d\log E_M/dt$ of the magnetic field. This suggests that the decay is constrained by the large-scale structure of the system rather than the microscopic physics and is in line with the results from \citet{ refId0} who reported only weak scaling of the decay timescale with Lundquist number. \citet{Reppin2017NonhelicalTA} have reported less inverse transfer at higher Lundquist numbers, while \citet{PhysRevE.107.055206} got the opposite result. Since the former study used $n=3$ hyper-diffusion while the latter used ordinary Laplacian diffusion, it might be the case that the order of the diffusion operator has a significant impact on this scaling. Resolving this will require further parameter studies which lie outside the scope of this paper. 

\section{Conclusions}\label{sec:conclusions}

In this paper, we analyzed the shell-to-shell energy and helicity transfer functions for simulations of decaying helical and non-helical MHD turbulence. The main results of our diagnostics are the following. 

\begin{itemize}
    \item For both helical and non-helical systems: 
    \begin{itemize}
        \item Energy is being transferred non-locally from the integral scale directly into large-scale (IR) magnetic modes. 
        \item  At the integral scale, both, the velocity and magnetic fields donate comparable amounts of energy to the IR-magnetic modes via advection as well as magnetic tension forces. The relative contribution of these channels depends on the total helicity of the system.
        \item This energy-transfer leads to a scale-independent multiplicative growth of the IR modes $\frac{1}{E_k}\frac{d E_k}{dt} = \frac{1}{\tau(t)}$, with a timescale $\tau\propto t$, consistent with a self-similar shift of the power-spectrum.  
        \item Upscale transfer of magnetic helicity coincides with upscale transfer within the magnetic energy reservoir. 
    \end{itemize}
    \item In the case of vanishing net helicity, energy and helicity transfer functions can be decomposed into positively and negatively helical modes. Inverse transfer is only found within each helical sector, not across them. This is consistent with the mutual annihilation of local structures with oppositely signed helicity, as discussed in HS. 
\end{itemize}

We identify energy and helicity shell-to-shell transfer functions as a promising tool for analyzing MHD turbulence beyond forced steady-state scenarios. In systems where helicity is dynamically important, helical mode decomposition provides a natural and useful extension of this framework. The present study is restricted to the nearly incompressible regime. Since the shell-to-shell formalism has been extended to compressible MHD \citep{10.1063/1.4990613}, applying similar diagnostics to compressible turbulence would be a natural next step. In this case, magnetic and fluid pressure introduces additional energy-transfer functions that can be dynamically important. We applied the helical-mode decomposition to the helicity and energy transfer functions in the magnetic reservoir to test the picture of helical mergers presented in HS. Helical-mode decomposition can also be applied to the velocity field, where, in the compressive case, an additional compressive mode is added. For future work, it would be interesting to apply a decomposition to all interacting shells in all transfer channels to find out whether the relative polarization of the interacting modes between the magnetic and velocity fields matters for inverse transfer. 

Inverse transfer in decaying MHD turbulence has a direct application for the evolution of primordial magnetic fields (PMFs) in the early universe \citep{PhysRevD.70.123003}. Here, the evolution of the magnetic energy-density $E$ and the correlation length $L$ are important, as they are required to constrain different generation mechanisms with present-day observables, e.g. the field strength in cosmic voids. It has been argued that non-helical inverse transfer controlled by the Hosking integral, whose underlying theory we have tested in this paper, can reconcile void fields with cosmic magnetogenesis \cite{Hosking:2022umv}. 

Another way we might be able to constrain PMF is via the cosmic microwave background (CMB). This is because PMFs would source density perturbations in the primordial plasma, which, if strong enough, would cause imprints in the CMB via spectral distortions \citep{2025PhLB..86539456U, 2015PhRvD..92l3004W}. Understanding this process will require a detailed analysis of the interaction between the magnetic and density fields on different scales, which is what the compressive shell-to-shell energy-transfer functions are designed to provide.

\begin{acknowledgments}
LK would like to thank Axel Brandenburg for helpful discussions during the NORDITA winter school on cosmological magnetic fields 2026.  
PG acknowledges funding by the Deutsche Forschungsgemeinschaft (DFG) – 555983577.
MB acknowledges funding by the Deutsche Forschungsgemeinschaft (DFG) under Germany's Excellence Strategy -- EXC 2121 ``Quantum Universe" --  390833306 and the DFG Research Group "Relativistic Jets" FOR5195 – project number 443220636. 
This project received access to JUPITER through the JUPITER Research and Early Access Program (JUREAP). JUPITER is funded by the EuroHPC Joint Undertaking, the German Federal Ministry of Research, Technology and Space, and the Ministry of Culture and Science of the German state of North Rhine-Westphalia.
\texttt{AthenaPK} and \texttt{Parthenon} use the \texttt{Kokkos} library \citep{9485033}. Analysis and plotting routines made use of the packages \texttt{yt} \citep{2011ApJS..192....9T}, \texttt{NumPy} \citep{harris2020array}, \texttt{scipy} \citep{2020SciPy-NMeth} and \texttt{Matplotlib} \citep{Hunter:2007}.
\end{acknowledgments}

\appendix

\section{Convergence}\label{sec:convergence}

\begin{figure}
    \centering
    \includegraphics[width=\linewidth]{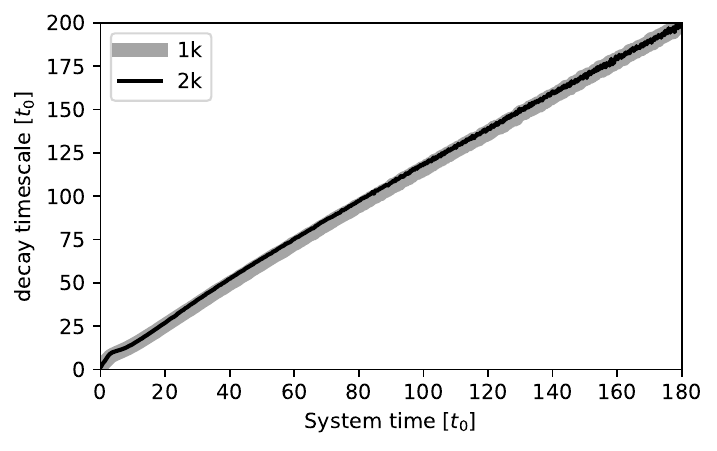}
    \caption{The magnetic energy decay timescale $\tau(t) = [(1/E_M) \:d E_M/dt]^{-1}$ is compared for the non-helical run parameters at $1024^3$ (1k) and $2048^3$ (2k) resolutions.}
    \label{fig:convergence_decay_timescales}
\end{figure}

\begin{figure}
    \centering
    \includegraphics[width=\linewidth]{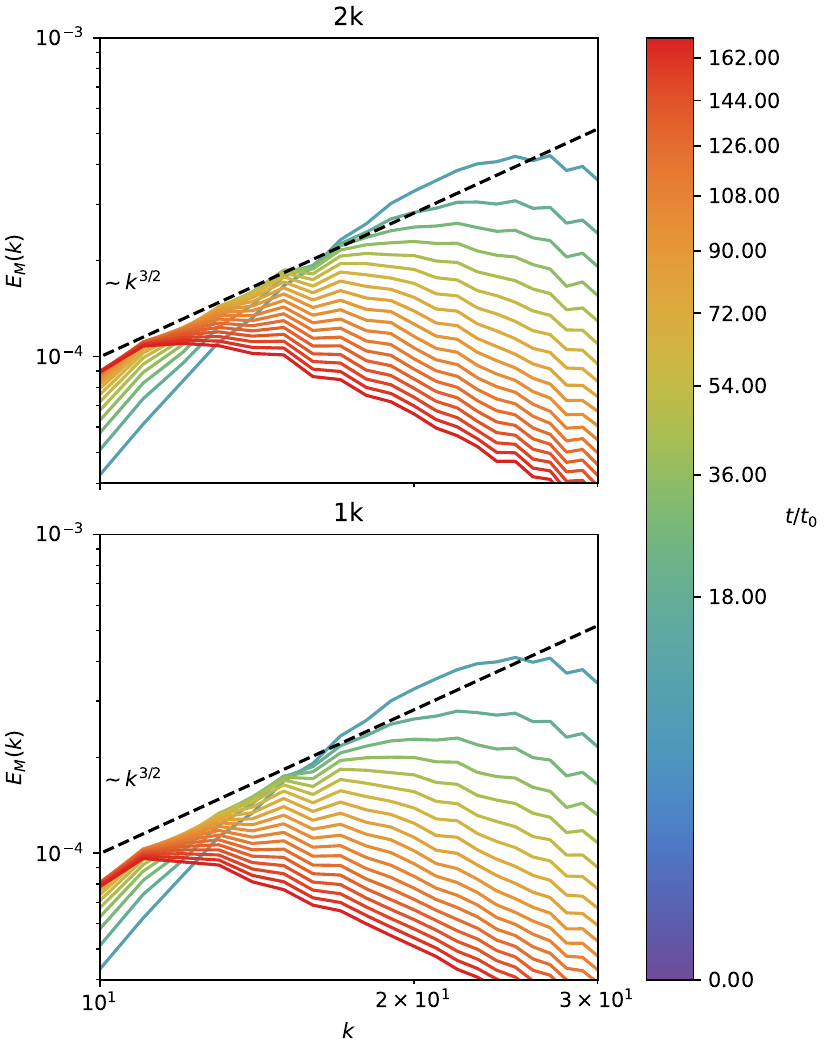}
    \caption{The magnetic energy power spectrum evolution is compared for the non-helical run parameters at $2048^3$ (2k) and $1024^3$ (1k) resolutions.}
    \label{fig:convergence_spectra}
\end{figure}

We tested the convergence of our simulation by running the non-helical setup with a $1024^3$ (1k) box and otherwise equal parameters. In Fig. \ref{fig:convergence_spectra}, we compare the evolution of the magnetic energy power spectra for both cases. We find that for 1k, the slope of the spectral peak is steeper than the $\sim k_I^{3/2}$ expected from the conservation of the Hosking integral, while good agreement with the expected trend is found with our production-run box size of $2048^3$ (2k). This suggests, that at 1k, the scale separation between integral and numerical diffusion scales is not large enough to call the system converged. We furthermore compare the total magnetic energy decay timescales $\tau(t) = [(1/E_M) \:d E_M/dt]^{-1}$ in figure \ref{fig:convergence_decay_timescales}. They are close to equal at all times, suggesting that the decay is independent of the numerical diffusion scale, provided the scale separation is large enough. 

\bibliographystyle{apsrev4-2}
\bibliography{sample701}

\begin{thebibliography}{53}%
\makeatletter
\providecommand \@ifxundefined [1]{%
 \@ifx{#1\undefined}
}%
\providecommand \@ifnum [1]{%
 \ifnum #1\expandafter \@firstoftwo
 \else \expandafter \@secondoftwo
 \fi
}%
\providecommand \@ifx [1]{%
 \ifx #1\expandafter \@firstoftwo
 \else \expandafter \@secondoftwo
 \fi
}%
\providecommand \natexlab [1]{#1}%
\providecommand \enquote  [1]{``#1''}%
\providecommand \bibnamefont  [1]{#1}%
\providecommand \bibfnamefont [1]{#1}%
\providecommand \citenamefont [1]{#1}%
\providecommand \href@noop [0]{\@secondoftwo}%
\providecommand \href [0]{\begingroup \@sanitize@url \@href}%
\providecommand \@href[1]{\@@startlink{#1}\@@href}%
\providecommand \@@href[1]{\endgroup#1\@@endlink}%
\providecommand \@sanitize@url [0]{\catcode `\\12\catcode `\$12\catcode `\&12\catcode `\#12\catcode `\^12\catcode `\_12\catcode `\%12\relax}%
\providecommand \@@startlink[1]{}%
\providecommand \@@endlink[0]{}%
\providecommand \url  [0]{\begingroup\@sanitize@url \@url }%
\providecommand \@url [1]{\endgroup\@href {#1}{\urlprefix }}%
\providecommand \urlprefix  [0]{URL }%
\providecommand \Eprint [0]{\href }%
\providecommand \doibase [0]{https://doi.org/}%
\providecommand \selectlanguage [0]{\@gobble}%
\providecommand \bibinfo  [0]{\@secondoftwo}%
\providecommand \bibfield  [0]{\@secondoftwo}%
\providecommand \translation [1]{[#1]}%
\providecommand \BibitemOpen [0]{}%
\providecommand \bibitemStop [0]{}%
\providecommand \bibitemNoStop [0]{.\EOS\space}%
\providecommand \EOS [0]{\spacefactor3000\relax}%
\providecommand \BibitemShut  [1]{\csname bibitem#1\endcsname}%
\let\auto@bib@innerbib\@empty
\bibitem [{\citenamefont {{Schekochihin}}(2022)}]{2022JPlPh..88e1501S}%
  \BibitemOpen
  \bibfield  {author} {\bibinfo {author} {\bibfnamefont {A.~A.}\ \bibnamefont {{Schekochihin}}},\ }\href {https://doi.org/10.1017/S0022377822000721} {\bibfield  {journal} {\bibinfo  {journal} {Journal of Plasma Physics}\ }\textbf {\bibinfo {volume} {88}},\ \bibinfo {eid} {155880501} (\bibinfo {year} {2022})},\ \Eprint {https://arxiv.org/abs/2010.00699} {arXiv:2010.00699 [physics.plasm-ph]} \BibitemShut {NoStop}%
\bibitem [{\citenamefont {Banerjee}\ and\ \citenamefont {Jedamzik}(2004)}]{PhysRevD.70.123003}%
  \BibitemOpen
  \bibfield  {author} {\bibinfo {author} {\bibfnamefont {R.}~\bibnamefont {Banerjee}}\ and\ \bibinfo {author} {\bibfnamefont {K.}~\bibnamefont {Jedamzik}},\ }\href {https://doi.org/10.1103/PhysRevD.70.123003} {\bibfield  {journal} {\bibinfo  {journal} {Phys. Rev. D}\ }\textbf {\bibinfo {volume} {70}},\ \bibinfo {pages} {123003} (\bibinfo {year} {2004})}\BibitemShut {NoStop}%
\bibitem [{\citenamefont {Durrer}\ and\ \citenamefont {Neronov}(2013)}]{Durrer:2013pga}%
  \BibitemOpen
  \bibfield  {author} {\bibinfo {author} {\bibfnamefont {R.}~\bibnamefont {Durrer}}\ and\ \bibinfo {author} {\bibfnamefont {A.}~\bibnamefont {Neronov}},\ }\href {https://doi.org/10.1007/s00159-013-0062-7} {\bibfield  {journal} {\bibinfo  {journal} {Astron. Astrophys. Rev.}\ }\textbf {\bibinfo {volume} {21}},\ \bibinfo {pages} {62} (\bibinfo {year} {2013})},\ \Eprint {https://arxiv.org/abs/1303.7121} {arXiv:1303.7121 [astro-ph.CO]} \BibitemShut {NoStop}%
\bibitem [{\citenamefont {Subramanian}(2016)}]{Subramanian_2016}%
  \BibitemOpen
  \bibfield  {author} {\bibinfo {author} {\bibfnamefont {K.}~\bibnamefont {Subramanian}},\ }\href {https://doi.org/10.1088/0034-4885/79/7/076901} {\bibfield  {journal} {\bibinfo  {journal} {Reports on Progress in Physics}\ }\textbf {\bibinfo {volume} {79}},\ \bibinfo {pages} {076901} (\bibinfo {year} {2016})}\BibitemShut {NoStop}%
\bibitem [{\citenamefont {Chen}\ \emph {et~al.}(2011)\citenamefont {Chen}, \citenamefont {Mallet}, \citenamefont {Yousef}, \citenamefont {Schekochihin},\ and\ \citenamefont {Horbury}}]{10.1111/j.1365-2966.2011.18933.x}%
  \BibitemOpen
  \bibfield  {author} {\bibinfo {author} {\bibfnamefont {C.~H.~K.}\ \bibnamefont {Chen}}, \bibinfo {author} {\bibfnamefont {A.}~\bibnamefont {Mallet}}, \bibinfo {author} {\bibfnamefont {T.~A.}\ \bibnamefont {Yousef}}, \bibinfo {author} {\bibfnamefont {A.~A.}\ \bibnamefont {Schekochihin}},\ and\ \bibinfo {author} {\bibfnamefont {T.~S.}\ \bibnamefont {Horbury}},\ }\href {https://doi.org/10.1111/j.1365-2966.2011.18933.x} {\bibfield  {journal} {\bibinfo  {journal} {Monthly Notices of the Royal Astronomical Society}\ }\textbf {\bibinfo {volume} {415}},\ \bibinfo {pages} {3219} (\bibinfo {year} {2011})},\ \Eprint {https://arxiv.org/abs/https://academic.oup.com/mnras/article-pdf/415/4/3219/4895690/mnras0415-3219.pdf} {https://academic.oup.com/mnras/article-pdf/415/4/3219/4895690/mnras0415-3219.pdf} \BibitemShut {NoStop}%
\bibitem [{\citenamefont {{Kolmogorov}}(1941)}]{1941DoSSR..30..301K}%
  \BibitemOpen
  \bibfield  {author} {\bibinfo {author} {\bibfnamefont {A.}~\bibnamefont {{Kolmogorov}}},\ }\href@noop {} {\bibfield  {journal} {\bibinfo  {journal} {Akademiia Nauk SSSR Doklady}\ }\textbf {\bibinfo {volume} {30}},\ \bibinfo {pages} {301} (\bibinfo {year} {1941})}\BibitemShut {NoStop}%
\bibitem [{\citenamefont {Woltjer}(1958)}]{doi:10.1073/pnas.44.6.489}%
  \BibitemOpen
  \bibfield  {author} {\bibinfo {author} {\bibfnamefont {L.}~\bibnamefont {Woltjer}},\ }\href {https://doi.org/10.1073/pnas.44.6.489} {\bibfield  {journal} {\bibinfo  {journal} {Proceedings of the National Academy of Sciences}\ }\textbf {\bibinfo {volume} {44}},\ \bibinfo {pages} {489} (\bibinfo {year} {1958})},\ \Eprint {https://arxiv.org/abs/https://www.pnas.org/doi/pdf/10.1073/pnas.44.6.489} {https://www.pnas.org/doi/pdf/10.1073/pnas.44.6.489} \BibitemShut {NoStop}%
\bibitem [{\citenamefont {Kahniashvili}\ \emph {et~al.}(2010)\citenamefont {Kahniashvili}, \citenamefont {Brandenburg}, \citenamefont {Tevzadze},\ and\ \citenamefont {Ratra}}]{PhysRevD.81.123002}%
  \BibitemOpen
  \bibfield  {author} {\bibinfo {author} {\bibfnamefont {T.}~\bibnamefont {Kahniashvili}}, \bibinfo {author} {\bibfnamefont {A.}~\bibnamefont {Brandenburg}}, \bibinfo {author} {\bibfnamefont {A.~G.}\ \bibnamefont {Tevzadze}},\ and\ \bibinfo {author} {\bibfnamefont {B.}~\bibnamefont {Ratra}},\ }\href {https://doi.org/10.1103/PhysRevD.81.123002} {\bibfield  {journal} {\bibinfo  {journal} {Phys. Rev. D}\ }\textbf {\bibinfo {volume} {81}},\ \bibinfo {pages} {123002} (\bibinfo {year} {2010})}\BibitemShut {NoStop}%
\bibitem [{\citenamefont {Pouquet}\ \emph {et~al.}(1976)\citenamefont {Pouquet}, \citenamefont {Frisch},\ and\ \citenamefont {Léorat}}]{Pouquet_Frisch_Léorat_1976}%
  \BibitemOpen
  \bibfield  {author} {\bibinfo {author} {\bibfnamefont {A.}~\bibnamefont {Pouquet}}, \bibinfo {author} {\bibfnamefont {U.}~\bibnamefont {Frisch}},\ and\ \bibinfo {author} {\bibfnamefont {J.}~\bibnamefont {Léorat}},\ }\href {https://doi.org/10.1017/S0022112076002140} {\bibfield  {journal} {\bibinfo  {journal} {Journal of Fluid Mechanics}\ }\textbf {\bibinfo {volume} {77}},\ \bibinfo {pages} {321–354} (\bibinfo {year} {1976})}\BibitemShut {NoStop}%
\bibitem [{\citenamefont {Christensson}\ \emph {et~al.}(2001)\citenamefont {Christensson}, \citenamefont {Hindmarsh},\ and\ \citenamefont {Brandenburg}}]{PhysRevE.64.056405}%
  \BibitemOpen
  \bibfield  {author} {\bibinfo {author} {\bibfnamefont {M.}~\bibnamefont {Christensson}}, \bibinfo {author} {\bibfnamefont {M.}~\bibnamefont {Hindmarsh}},\ and\ \bibinfo {author} {\bibfnamefont {A.}~\bibnamefont {Brandenburg}},\ }\href {https://doi.org/10.1103/PhysRevE.64.056405} {\bibfield  {journal} {\bibinfo  {journal} {Phys. Rev. E}\ }\textbf {\bibinfo {volume} {64}},\ \bibinfo {pages} {056405} (\bibinfo {year} {2001})}\BibitemShut {NoStop}%
\bibitem [{\citenamefont {M\"uller}\ \emph {et~al.}(2012)\citenamefont {M\"uller}, \citenamefont {Malapaka},\ and\ \citenamefont {Busse}}]{PhysRevE.85.015302}%
  \BibitemOpen
  \bibfield  {author} {\bibinfo {author} {\bibfnamefont {W.-C.}\ \bibnamefont {M\"uller}}, \bibinfo {author} {\bibfnamefont {S.~K.}\ \bibnamefont {Malapaka}},\ and\ \bibinfo {author} {\bibfnamefont {A.}~\bibnamefont {Busse}},\ }\href {https://doi.org/10.1103/PhysRevE.85.015302} {\bibfield  {journal} {\bibinfo  {journal} {Phys. Rev. E}\ }\textbf {\bibinfo {volume} {85}},\ \bibinfo {pages} {015302} (\bibinfo {year} {2012})}\BibitemShut {NoStop}%
\bibitem [{\citenamefont {Brandenburg}\ \emph {et~al.}(2015)\citenamefont {Brandenburg}, \citenamefont {Kahniashvili},\ and\ \citenamefont {Tevzadze}}]{PhysRevLett.114.075001}%
  \BibitemOpen
  \bibfield  {author} {\bibinfo {author} {\bibfnamefont {A.}~\bibnamefont {Brandenburg}}, \bibinfo {author} {\bibfnamefont {T.}~\bibnamefont {Kahniashvili}},\ and\ \bibinfo {author} {\bibfnamefont {A.~G.}\ \bibnamefont {Tevzadze}},\ }\href {https://doi.org/10.1103/PhysRevLett.114.075001} {\bibfield  {journal} {\bibinfo  {journal} {Phys. Rev. Lett.}\ }\textbf {\bibinfo {volume} {114}},\ \bibinfo {pages} {075001} (\bibinfo {year} {2015})}\BibitemShut {NoStop}%
\bibitem [{\citenamefont {Hosking}\ and\ \citenamefont {Schekochihin}(2021)}]{PhysRevX.11.041005}%
  \BibitemOpen
  \bibfield  {author} {\bibinfo {author} {\bibfnamefont {D.~N.}\ \bibnamefont {Hosking}}\ and\ \bibinfo {author} {\bibfnamefont {A.~A.}\ \bibnamefont {Schekochihin}},\ }\href {https://doi.org/10.1103/PhysRevX.11.041005} {\bibfield  {journal} {\bibinfo  {journal} {Phys. Rev. X}\ }\textbf {\bibinfo {volume} {11}},\ \bibinfo {pages} {041005} (\bibinfo {year} {2021})}\BibitemShut {NoStop}%
\bibitem [{\citenamefont {{Brandenburg}}\ \emph {et~al.}(2023)\citenamefont {{Brandenburg}}, \citenamefont {{Sharma}},\ and\ \citenamefont {{Vachaspati}}}]{2023JPlPh..89f9006B}%
  \BibitemOpen
  \bibfield  {author} {\bibinfo {author} {\bibfnamefont {A.}~\bibnamefont {{Brandenburg}}}, \bibinfo {author} {\bibfnamefont {R.}~\bibnamefont {{Sharma}}},\ and\ \bibinfo {author} {\bibfnamefont {T.}~\bibnamefont {{Vachaspati}}},\ }\href {https://doi.org/10.1017/S0022377823001253} {\bibfield  {journal} {\bibinfo  {journal} {Journal of Plasma Physics}\ }\textbf {\bibinfo {volume} {89}},\ \bibinfo {eid} {905890606} (\bibinfo {year} {2023})},\ \Eprint {https://arxiv.org/abs/2307.04602} {arXiv:2307.04602 [physics.plasm-ph]} \BibitemShut {NoStop}%
\bibitem [{Note1()}]{Note1}%
  \BibitemOpen
  \bibinfo {note} {For spectra with a shallower subinertial range, the dynamics may instead be controlled by other conserved quantities, such as the Saffman integral in the $k^2$ case \cite {2023JPlPh..89f9006B}.}\BibitemShut {Stop}%
\bibitem [{\citenamefont {Waleffe}(1992)}]{Waleffe1992}%
  \BibitemOpen
  \bibfield  {author} {\bibinfo {author} {\bibfnamefont {F.}~\bibnamefont {Waleffe}},\ }\href {https://doi.org/10.1063/1.858309} {\bibfield  {journal} {\bibinfo  {journal} {Physics of Fluids A: Fluid Dynamics}\ }\textbf {\bibinfo {volume} {4}},\ \bibinfo {pages} {350} (\bibinfo {year} {1992})}\BibitemShut {NoStop}%
\bibitem [{\citenamefont {Linkmann}\ \emph {et~al.}(2016)\citenamefont {Linkmann}, \citenamefont {Berera}, \citenamefont {McKay},\ and\ \citenamefont {Jäger}}]{Linkmann_Berera_McKay_Jäger_2016}%
  \BibitemOpen
  \bibfield  {author} {\bibinfo {author} {\bibfnamefont {M.}~\bibnamefont {Linkmann}}, \bibinfo {author} {\bibfnamefont {A.}~\bibnamefont {Berera}}, \bibinfo {author} {\bibfnamefont {M.}~\bibnamefont {McKay}},\ and\ \bibinfo {author} {\bibfnamefont {J.}~\bibnamefont {Jäger}},\ }\href {https://doi.org/10.1017/jfm.2016.43} {\bibfield  {journal} {\bibinfo  {journal} {Journal of Fluid Mechanics}\ }\textbf {\bibinfo {volume} {791}},\ \bibinfo {pages} {61–96} (\bibinfo {year} {2016})}\BibitemShut {NoStop}%
\bibitem [{\citenamefont {Alexakis}\ \emph {et~al.}(2005)\citenamefont {Alexakis}, \citenamefont {Mininni},\ and\ \citenamefont {Pouquet}}]{PhysRevE.72.046301}%
  \BibitemOpen
  \bibfield  {author} {\bibinfo {author} {\bibfnamefont {A.}~\bibnamefont {Alexakis}}, \bibinfo {author} {\bibfnamefont {P.~D.}\ \bibnamefont {Mininni}},\ and\ \bibinfo {author} {\bibfnamefont {A.}~\bibnamefont {Pouquet}},\ }\href {https://doi.org/10.1103/PhysRevE.72.046301} {\bibfield  {journal} {\bibinfo  {journal} {Phys. Rev. E}\ }\textbf {\bibinfo {volume} {72}},\ \bibinfo {pages} {046301} (\bibinfo {year} {2005})}\BibitemShut {NoStop}%
\bibitem [{\citenamefont {Dar}\ \emph {et~al.}(2001)\citenamefont {Dar}, \citenamefont {Verma},\ and\ \citenamefont {Eswaran}}]{DAR2001207}%
  \BibitemOpen
  \bibfield  {author} {\bibinfo {author} {\bibfnamefont {G.}~\bibnamefont {Dar}}, \bibinfo {author} {\bibfnamefont {M.~K.}\ \bibnamefont {Verma}},\ and\ \bibinfo {author} {\bibfnamefont {V.}~\bibnamefont {Eswaran}},\ }\href {https://doi.org/https://doi.org/10.1016/S0167-2789(01)00307-4} {\bibfield  {journal} {\bibinfo  {journal} {Physica D: Nonlinear Phenomena}\ }\textbf {\bibinfo {volume} {157}},\ \bibinfo {pages} {207} (\bibinfo {year} {2001})}\BibitemShut {NoStop}%
\bibitem [{\citenamefont {{Teaca}}\ \emph {et~al.}(2011)\citenamefont {{Teaca}}, \citenamefont {{Carati}},\ and\ \citenamefont {{Andrzej Domaradzki}}}]{2011PhPl...18k2307T}%
  \BibitemOpen
  \bibfield  {author} {\bibinfo {author} {\bibfnamefont {B.}~\bibnamefont {{Teaca}}}, \bibinfo {author} {\bibfnamefont {D.}~\bibnamefont {{Carati}}},\ and\ \bibinfo {author} {\bibfnamefont {J.}~\bibnamefont {{Andrzej Domaradzki}}},\ }\href {https://doi.org/10.1063/1.3661086} {\bibfield  {journal} {\bibinfo  {journal} {Physics of Plasmas}\ }\textbf {\bibinfo {volume} {18}},\ \bibinfo {eid} {112307} (\bibinfo {year} {2011})},\ \Eprint {https://arxiv.org/abs/1108.3937} {arXiv:1108.3937 [physics.flu-dyn]} \BibitemShut {NoStop}%
\bibitem [{\citenamefont {Grete}\ \emph {et~al.}(2017)\citenamefont {Grete}, \citenamefont {O'Shea}, \citenamefont {Beckwith}, \citenamefont {Schmidt},\ and\ \citenamefont {Christlieb}}]{10.1063/1.4990613}%
  \BibitemOpen
  \bibfield  {author} {\bibinfo {author} {\bibfnamefont {P.}~\bibnamefont {Grete}}, \bibinfo {author} {\bibfnamefont {B.~W.}\ \bibnamefont {O'Shea}}, \bibinfo {author} {\bibfnamefont {K.}~\bibnamefont {Beckwith}}, \bibinfo {author} {\bibfnamefont {W.}~\bibnamefont {Schmidt}},\ and\ \bibinfo {author} {\bibfnamefont {A.}~\bibnamefont {Christlieb}},\ }\href {https://doi.org/10.1063/1.4990613} {\bibfield  {journal} {\bibinfo  {journal} {Physics of Plasmas}\ }\textbf {\bibinfo {volume} {24}},\ \bibinfo {pages} {092311} (\bibinfo {year} {2017})}\BibitemShut {NoStop}%
\bibitem [{\citenamefont {Hew}\ \emph {et~al.}(2026)\citenamefont {Hew}, \citenamefont {Hosking}, \citenamefont {Federrath}, \citenamefont {Beattie},\ and\ \citenamefont {Kriel}}]{hew_hosking_federrath_beattie_kriel_2026}%
  \BibitemOpen
  \bibfield  {author} {\bibinfo {author} {\bibfnamefont {J.~K.~J.}\ \bibnamefont {Hew}}, \bibinfo {author} {\bibfnamefont {D.~N.}\ \bibnamefont {Hosking}}, \bibinfo {author} {\bibfnamefont {C.}~\bibnamefont {Federrath}}, \bibinfo {author} {\bibfnamefont {J.~R.}\ \bibnamefont {Beattie}},\ and\ \bibinfo {author} {\bibfnamefont {N.}~\bibnamefont {Kriel}},\ }\bibfield  {journal} {\bibinfo  {journal} {Journal of Plasma Physics}\ }\textbf {\bibinfo {volume} {92}},\ \href {https://doi.org/10.1017/s0022377826101275} {10.1017/s0022377826101275} (\bibinfo {year} {2026})\BibitemShut {NoStop}%
\bibitem [{\citenamefont {Subramanian}\ and\ \citenamefont {Brandenburg}(2006)}]{Subramanian_2006}%
  \BibitemOpen
  \bibfield  {author} {\bibinfo {author} {\bibfnamefont {K.}~\bibnamefont {Subramanian}}\ and\ \bibinfo {author} {\bibfnamefont {A.}~\bibnamefont {Brandenburg}},\ }\href {https://doi.org/10.1086/507828} {\bibfield  {journal} {\bibinfo  {journal} {The Astrophysical Journal}\ }\textbf {\bibinfo {volume} {648}},\ \bibinfo {pages} {L71} (\bibinfo {year} {2006})}\BibitemShut {NoStop}%
\bibitem [{Note2()}]{Note2}%
  \BibitemOpen
  \bibinfo {note} {\protect \texttt {AthenaPK} is available and maintained at https://github.com/parthenon-hpc-lab/athenapk and commit ca99dd4 was used for the simulations.}\BibitemShut {Stop}%
\bibitem [{\citenamefont {Grete}\ \emph {et~al.}(2023{\natexlab{a}})\citenamefont {Grete}, \citenamefont {Dolence}, \citenamefont {Miller}, \citenamefont {Brown}, \citenamefont {Ryan}, \citenamefont {Gaspar}, \citenamefont {Glines}, \citenamefont {Swaminarayan}, \citenamefont {Lippuner}, \citenamefont {Solomon}, \citenamefont {Shipman}, \citenamefont {Junghans}, \citenamefont {Holladay}, \citenamefont {Stone},\ and\ \citenamefont {Roberts}}]{parthenon}%
  \BibitemOpen
  \bibfield  {author} {\bibinfo {author} {\bibfnamefont {P.}~\bibnamefont {Grete}}, \bibinfo {author} {\bibfnamefont {J.~C.}\ \bibnamefont {Dolence}}, \bibinfo {author} {\bibfnamefont {J.~M.}\ \bibnamefont {Miller}}, \bibinfo {author} {\bibfnamefont {J.}~\bibnamefont {Brown}}, \bibinfo {author} {\bibfnamefont {B.}~\bibnamefont {Ryan}}, \bibinfo {author} {\bibfnamefont {A.}~\bibnamefont {Gaspar}}, \bibinfo {author} {\bibfnamefont {F.}~\bibnamefont {Glines}}, \bibinfo {author} {\bibfnamefont {S.}~\bibnamefont {Swaminarayan}}, \bibinfo {author} {\bibfnamefont {J.}~\bibnamefont {Lippuner}}, \bibinfo {author} {\bibfnamefont {C.~J.}\ \bibnamefont {Solomon}}, \bibinfo {author} {\bibfnamefont {G.}~\bibnamefont {Shipman}}, \bibinfo {author} {\bibfnamefont {C.}~\bibnamefont {Junghans}}, \bibinfo {author} {\bibfnamefont {D.}~\bibnamefont {Holladay}}, \bibinfo {author} {\bibfnamefont {J.~M.}\ \bibnamefont {Stone}},\ and\ \bibinfo {author} {\bibfnamefont {L.~F.}\ \bibnamefont {Roberts}},\ }\href
  {https://doi.org/10.1177/10943420221143775} {\bibfield  {journal} {\bibinfo  {journal} {The International Journal of High Performance Computing Applications}\ }\textbf {\bibinfo {volume} {37}},\ \bibinfo {pages} {465} (\bibinfo {year} {2023}{\natexlab{a}})}\BibitemShut {NoStop}%
\bibitem [{\citenamefont {Miyoshi}\ and\ \citenamefont {Kusano}(2005)}]{HLLD}%
  \BibitemOpen
  \bibfield  {author} {\bibinfo {author} {\bibfnamefont {T.}~\bibnamefont {Miyoshi}}\ and\ \bibinfo {author} {\bibfnamefont {K.}~\bibnamefont {Kusano}},\ }\href {https://doi.org/10.1016/j.jcp.2005.02.017} {\bibfield  {journal} {\bibinfo  {journal} {Journal of Computational Physics}\ }\textbf {\bibinfo {volume} {208}},\ \bibinfo {pages} {315 } (\bibinfo {year} {2005})}\BibitemShut {NoStop}%
\bibitem [{\citenamefont {Dedner}\ \emph {et~al.}(2002)\citenamefont {Dedner}, \citenamefont {Kemm}, \citenamefont {Kröner}, \citenamefont {Munz}, \citenamefont {Schnitzer},\ and\ \citenamefont {Wesenberg}}]{dedner2002}%
  \BibitemOpen
  \bibfield  {author} {\bibinfo {author} {\bibfnamefont {A.}~\bibnamefont {Dedner}}, \bibinfo {author} {\bibfnamefont {F.}~\bibnamefont {Kemm}}, \bibinfo {author} {\bibfnamefont {D.}~\bibnamefont {Kröner}}, \bibinfo {author} {\bibfnamefont {C.-D.}\ \bibnamefont {Munz}}, \bibinfo {author} {\bibfnamefont {T.}~\bibnamefont {Schnitzer}},\ and\ \bibinfo {author} {\bibfnamefont {M.}~\bibnamefont {Wesenberg}},\ }\href {https://doi.org/10.1006/jcph.2001.6961} {\bibfield  {journal} {\bibinfo  {journal} {Journal of Computational Physics}\ }\textbf {\bibinfo {volume} {175}},\ \bibinfo {pages} {645 } (\bibinfo {year} {2002})}\BibitemShut {NoStop}%
\bibitem [{\citenamefont {Grete}\ \emph {et~al.}(2023{\natexlab{b}})\citenamefont {Grete}, \citenamefont {O’Shea},\ and\ \citenamefont {Beckwith}}]{Grete2023}%
  \BibitemOpen
  \bibfield  {author} {\bibinfo {author} {\bibfnamefont {P.}~\bibnamefont {Grete}}, \bibinfo {author} {\bibfnamefont {B.~W.}\ \bibnamefont {O’Shea}},\ and\ \bibinfo {author} {\bibfnamefont {K.}~\bibnamefont {Beckwith}},\ }\href {https://doi.org/10.3847/2041-8213/acaea7} {\bibfield  {journal} {\bibinfo  {journal} {The Astrophysical Journal Letters}\ }\textbf {\bibinfo {volume} {942}},\ \bibinfo {pages} {L34} (\bibinfo {year} {2023}{\natexlab{b}})}\BibitemShut {NoStop}%
\bibitem [{\citenamefont {Ayala}\ \emph {et~al.}(2020)\citenamefont {Ayala}, \citenamefont {Tomov}, \citenamefont {Haidar},\ and\ \citenamefont {Dongarra}}]{Ayala2020}%
  \BibitemOpen
  \bibfield  {author} {\bibinfo {author} {\bibfnamefont {A.}~\bibnamefont {Ayala}}, \bibinfo {author} {\bibfnamefont {S.}~\bibnamefont {Tomov}}, \bibinfo {author} {\bibfnamefont {A.}~\bibnamefont {Haidar}},\ and\ \bibinfo {author} {\bibfnamefont {J.}~\bibnamefont {Dongarra}},\ }in\ \href@noop {} {\emph {\bibinfo {booktitle} {Computational Science -- ICCS 2020}}},\ \bibinfo {editor} {edited by\ \bibinfo {editor} {\bibfnamefont {V.~V.}\ \bibnamefont {Krzhizhanovskaya}}, \bibinfo {editor} {\bibfnamefont {G.}~\bibnamefont {Z{\'a}vodszky}}, \bibinfo {editor} {\bibfnamefont {M.~H.}\ \bibnamefont {Lees}}, \bibinfo {editor} {\bibfnamefont {J.~J.}\ \bibnamefont {Dongarra}}, \bibinfo {editor} {\bibfnamefont {P.~M.~A.}\ \bibnamefont {Sloot}}, \bibinfo {editor} {\bibfnamefont {S.}~\bibnamefont {Brissos}},\ and\ \bibinfo {editor} {\bibfnamefont {J.}~\bibnamefont {Teixeira}}}\ (\bibinfo  {publisher} {Springer International Publishing},\ \bibinfo {address} {Cham},\ \bibinfo {year} {2020})\ pp.\ \bibinfo {pages}
  {262--275}\BibitemShut {NoStop}%
\bibitem [{\citenamefont {{Brandenburg, A.}}\ \emph {et~al.}(2024)\citenamefont {{Brandenburg, A.}}, \citenamefont {{Neronov, A.}},\ and\ \citenamefont {{Vazza, F.}}}]{refId0}%
  \BibitemOpen
  \bibfield  {author} {\bibinfo {author} {\bibnamefont {{Brandenburg, A.}}}, \bibinfo {author} {\bibnamefont {{Neronov, A.}}},\ and\ \bibinfo {author} {\bibnamefont {{Vazza, F.}}},\ }\href {https://doi.org/10.1051/0004-6361/202449267} {\bibfield  {journal} {\bibinfo  {journal} {A\&A}\ }\textbf {\bibinfo {volume} {687}},\ \bibinfo {pages} {A186} (\bibinfo {year} {2024})}\BibitemShut {NoStop}%
\bibitem [{\citenamefont {Reppin}\ and\ \citenamefont {Banerjee}(2017)}]{Reppin2017NonhelicalTA}%
  \BibitemOpen
  \bibfield  {author} {\bibinfo {author} {\bibfnamefont {J.}~\bibnamefont {Reppin}}\ and\ \bibinfo {author} {\bibfnamefont {R.}~\bibnamefont {Banerjee}},\ }\href {https://api.semanticscholar.org/CorpusID:25330345} {\bibfield  {journal} {\bibinfo  {journal} {Physical review. E}\ }\textbf {\bibinfo {volume} {96 5-1}},\ \bibinfo {pages} {053105} (\bibinfo {year} {2017})}\BibitemShut {NoStop}%
\bibitem [{Note3()}]{Note3}%
  \BibitemOpen
  \bibinfo {note} {The framework implements the formalism described in \protect \citet {10.1063/1.4990613}. It is available at \protect {\protect \url {https://github.com/pgrete/energy-transfer-analysis}} and makes use of the \protect \texttt {mpi4py-fft} library \protect \citep {mpi4py-fft}.}\BibitemShut {Stop}%
\bibitem [{\citenamefont {Aluie}\ and\ \citenamefont {Eyink}(2009)}]{Aluie2009}%
  \BibitemOpen
  \bibfield  {author} {\bibinfo {author} {\bibfnamefont {H.}~\bibnamefont {Aluie}}\ and\ \bibinfo {author} {\bibfnamefont {G.~L.}\ \bibnamefont {Eyink}},\ }\href {https://doi.org/10.1063/1.3266948} {\bibfield  {journal} {\bibinfo  {journal} {Physics of Fluids}\ }\textbf {\bibinfo {volume} {21}},\ \bibinfo {pages} {115108} (\bibinfo {year} {2009})}\BibitemShut {NoStop}%
\bibitem [{\citenamefont {Alexakis}\ \emph {et~al.}(2006)\citenamefont {Alexakis}, \citenamefont {Mininni},\ and\ \citenamefont {Pouquet}}]{Alexakis_2006}%
  \BibitemOpen
  \bibfield  {author} {\bibinfo {author} {\bibfnamefont {A.}~\bibnamefont {Alexakis}}, \bibinfo {author} {\bibfnamefont {P.~D.}\ \bibnamefont {Mininni}},\ and\ \bibinfo {author} {\bibfnamefont {A.}~\bibnamefont {Pouquet}},\ }\href {https://doi.org/10.1086/500082} {\bibfield  {journal} {\bibinfo  {journal} {The Astrophysical Journal}\ }\textbf {\bibinfo {volume} {640}},\ \bibinfo {pages} {335} (\bibinfo {year} {2006})}\BibitemShut {NoStop}%
\bibitem [{\citenamefont {Plunian}\ \emph {et~al.}(2019)\citenamefont {Plunian}, \citenamefont {Stepanov},\ and\ \citenamefont {Verma}}]{Plunian_Stepanov_Verma_2019}%
  \BibitemOpen
  \bibfield  {author} {\bibinfo {author} {\bibfnamefont {F.}~\bibnamefont {Plunian}}, \bibinfo {author} {\bibfnamefont {R.}~\bibnamefont {Stepanov}},\ and\ \bibinfo {author} {\bibfnamefont {M.~K.}\ \bibnamefont {Verma}},\ }\href {https://doi.org/10.1017/S0022377819000710} {\bibfield  {journal} {\bibinfo  {journal} {Journal of Plasma Physics}\ }\textbf {\bibinfo {volume} {85}},\ \bibinfo {pages} {905850507} (\bibinfo {year} {2019})}\BibitemShut {NoStop}%
\bibitem [{\citenamefont {Teissier}\ and\ \citenamefont {Müller}(2021)}]{Teissier_Müller_2021}%
  \BibitemOpen
  \bibfield  {author} {\bibinfo {author} {\bibfnamefont {J.-M.}\ \bibnamefont {Teissier}}\ and\ \bibinfo {author} {\bibfnamefont {W.-C.}\ \bibnamefont {Müller}},\ }\href {https://doi.org/10.1017/jfm.2021.496} {\bibfield  {journal} {\bibinfo  {journal} {Journal of Fluid Mechanics}\ }\textbf {\bibinfo {volume} {921}},\ \bibinfo {pages} {A7} (\bibinfo {year} {2021})}\BibitemShut {NoStop}%
\bibitem [{\citenamefont {Brandenburg}\ and\ \citenamefont {Larsson}(2023)}]{atmos14060932}%
  \BibitemOpen
  \bibfield  {author} {\bibinfo {author} {\bibfnamefont {A.}~\bibnamefont {Brandenburg}}\ and\ \bibinfo {author} {\bibfnamefont {G.}~\bibnamefont {Larsson}},\ }\bibfield  {journal} {\bibinfo  {journal} {Atmosphere}\ }\textbf {\bibinfo {volume} {14}},\ \href {https://doi.org/10.3390/atmos14060932} {10.3390/atmos14060932} (\bibinfo {year} {2023})\BibitemShut {NoStop}%
\bibitem [{Note4()}]{Note4}%
  \BibitemOpen
  \bibinfo {note} {Since $T_\protect \textnormal {BB}(Q,K)$ is asymmetric in $Q$ and $K$, the values at the diagonal are expected to be zero. Non-zero values across the diagonal are thus due to numerical errors, which are primarily caused by taking the gradients in equation \ref {eq:energy_transfer} in real space on a discrete mesh. They therefore only meaningfully affect the high-$k$ modes.}\BibitemShut {Stop}%
\bibitem [{\citenamefont {Grete}\ \emph {et~al.}(2021)\citenamefont {Grete}, \citenamefont {O’Shea},\ and\ \citenamefont {Beckwith}}]{Grete2021tension}%
  \BibitemOpen
  \bibfield  {author} {\bibinfo {author} {\bibfnamefont {P.}~\bibnamefont {Grete}}, \bibinfo {author} {\bibfnamefont {B.~W.}\ \bibnamefont {O’Shea}},\ and\ \bibinfo {author} {\bibfnamefont {K.}~\bibnamefont {Beckwith}},\ }\href {https://doi.org/10.3847/1538-4357/abdd22} {\bibfield  {journal} {\bibinfo  {journal} {The Astrophysical Journal}\ }\textbf {\bibinfo {volume} {909}},\ \bibinfo {pages} {148} (\bibinfo {year} {2021})}\BibitemShut {NoStop}%
\bibitem [{Note5()}]{Note5}%
  \BibitemOpen
  \bibinfo {note} {Note that even for the initially fully helical setup, helicity is no longer fully saturated at large $k$ at later times, as can be seen in Fig. \ref {fig:spectra}, explaining why a very small amount of direct helicity-transfer is still possible.}\BibitemShut {Stop}%
\bibitem [{\citenamefont {Bhat}\ \emph {et~al.}(2020)\citenamefont {Bhat}, \citenamefont {Zhou},\ and\ \citenamefont {Loureiro}}]{10.1093/mnras/staa3849}%
  \BibitemOpen
  \bibfield  {author} {\bibinfo {author} {\bibfnamefont {P.}~\bibnamefont {Bhat}}, \bibinfo {author} {\bibfnamefont {M.}~\bibnamefont {Zhou}},\ and\ \bibinfo {author} {\bibfnamefont {N.~F.}\ \bibnamefont {Loureiro}},\ }\href {https://doi.org/10.1093/mnras/staa3849} {\bibfield  {journal} {\bibinfo  {journal} {Monthly Notices of the Royal Astronomical Society}\ }\textbf {\bibinfo {volume} {501}},\ \bibinfo {pages} {3074} (\bibinfo {year} {2020})},\ \Eprint {https://arxiv.org/abs/https://academic.oup.com/mnras/article-pdf/501/2/3074/35573300/staa3849.pdf} {https://academic.oup.com/mnras/article-pdf/501/2/3074/35573300/staa3849.pdf} \BibitemShut {NoStop}%
\bibitem [{\citenamefont {Durrer}\ and\ \citenamefont {Caprini}(2003)}]{Ruth_Durrer_2003}%
  \BibitemOpen
  \bibfield  {author} {\bibinfo {author} {\bibfnamefont {R.}~\bibnamefont {Durrer}}\ and\ \bibinfo {author} {\bibfnamefont {C.}~\bibnamefont {Caprini}},\ }\href {https://doi.org/10.1088/1475-7516/2003/11/010} {\bibfield  {journal} {\bibinfo  {journal} {Journal of Cosmology and Astroparticle Physics}\ }\textbf {\bibinfo {volume} {2003}}\bibinfo  {number} { (11)},\ \bibinfo {pages} {010}}\BibitemShut {NoStop}%
\bibitem [{\citenamefont {Vachaspati}(2021)}]{Vachaspati_2021}%
  \BibitemOpen
\bibfield  {number} {  }\bibfield  {author} {\bibinfo {author} {\bibfnamefont {T.}~\bibnamefont {Vachaspati}},\ }\href {https://doi.org/10.1088/1361-6633/ac03a9} {\bibfield  {journal} {\bibinfo  {journal} {Reports on Progress in Physics}\ }\textbf {\bibinfo {volume} {84}},\ \bibinfo {pages} {074901} (\bibinfo {year} {2021})}\BibitemShut {NoStop}%
\bibitem [{\citenamefont {Armua}\ \emph {et~al.}(2023)\citenamefont {Armua}, \citenamefont {Berera},\ and\ \citenamefont {Calder\'on-Figueroa}}]{PhysRevE.107.055206}%
  \BibitemOpen
  \bibfield  {author} {\bibinfo {author} {\bibfnamefont {A.}~\bibnamefont {Armua}}, \bibinfo {author} {\bibfnamefont {A.}~\bibnamefont {Berera}},\ and\ \bibinfo {author} {\bibfnamefont {J.}~\bibnamefont {Calder\'on-Figueroa}},\ }\href {https://doi.org/10.1103/PhysRevE.107.055206} {\bibfield  {journal} {\bibinfo  {journal} {Phys. Rev. E}\ }\textbf {\bibinfo {volume} {107}},\ \bibinfo {pages} {055206} (\bibinfo {year} {2023})}\BibitemShut {NoStop}%
\bibitem [{\citenamefont {Hosking}\ and\ \citenamefont {Schekochihin}(2023)}]{Hosking:2022umv}%
  \BibitemOpen
  \bibfield  {author} {\bibinfo {author} {\bibfnamefont {D.~N.}\ \bibnamefont {Hosking}}\ and\ \bibinfo {author} {\bibfnamefont {A.~A.}\ \bibnamefont {Schekochihin}},\ }\href {https://doi.org/10.1038/s41467-023-43258-3} {\bibfield  {journal} {\bibinfo  {journal} {Nature Commun.}\ }\textbf {\bibinfo {volume} {14}},\ \bibinfo {pages} {7523} (\bibinfo {year} {2023})},\ \Eprint {https://arxiv.org/abs/2203.03573} {arXiv:2203.03573 [astro-ph.CO]} \BibitemShut {NoStop}%
\bibitem [{\citenamefont {{Uchida}}\ \emph {et~al.}(2025)\citenamefont {{Uchida}}, \citenamefont {{Kamada}},\ and\ \citenamefont {{Tashiro}}}]{2025PhLB..86539456U}%
  \BibitemOpen
  \bibfield  {author} {\bibinfo {author} {\bibfnamefont {F.}~\bibnamefont {{Uchida}}}, \bibinfo {author} {\bibfnamefont {K.}~\bibnamefont {{Kamada}}},\ and\ \bibinfo {author} {\bibfnamefont {H.}~\bibnamefont {{Tashiro}}},\ }\href {https://doi.org/10.1016/j.physletb.2025.139456} {\bibfield  {journal} {\bibinfo  {journal} {Physics Letters B}\ }\textbf {\bibinfo {volume} {865}},\ \bibinfo {eid} {139456} (\bibinfo {year} {2025})},\ \Eprint {https://arxiv.org/abs/2411.03183} {arXiv:2411.03183 [astro-ph.CO]} \BibitemShut {NoStop}%
\bibitem [{\citenamefont {{Wagstaff}}\ and\ \citenamefont {{Banerjee}}(2015)}]{2015PhRvD..92l3004W}%
  \BibitemOpen
  \bibfield  {author} {\bibinfo {author} {\bibfnamefont {J.~M.}\ \bibnamefont {{Wagstaff}}}\ and\ \bibinfo {author} {\bibfnamefont {R.}~\bibnamefont {{Banerjee}}},\ }\href {https://doi.org/10.1103/PhysRevD.92.123004} {\bibfield  {journal} {\bibinfo  {journal} {\prd}\ }\textbf {\bibinfo {volume} {92}},\ \bibinfo {eid} {123004} (\bibinfo {year} {2015})},\ \Eprint {https://arxiv.org/abs/1508.01683} {arXiv:1508.01683 [astro-ph.CO]} \BibitemShut {NoStop}%
\bibitem [{\citenamefont {Trott}\ \emph {et~al.}(2022)\citenamefont {Trott}, \citenamefont {Lebrun-Grandié}, \citenamefont {Arndt}, \citenamefont {Ciesko}, \citenamefont {Dang}, \citenamefont {Ellingwood}, \citenamefont {Gayatri}, \citenamefont {Harvey}, \citenamefont {Hollman}, \citenamefont {Ibanez}, \citenamefont {Liber}, \citenamefont {Madsen}, \citenamefont {Miles}, \citenamefont {Poliakoff}, \citenamefont {Powell}, \citenamefont {Rajamanickam}, \citenamefont {Simberg}, \citenamefont {Sunderland}, \citenamefont {Turcksin},\ and\ \citenamefont {Wilke}}]{9485033}%
  \BibitemOpen
  \bibfield  {author} {\bibinfo {author} {\bibfnamefont {C.~R.}\ \bibnamefont {Trott}}, \bibinfo {author} {\bibfnamefont {D.}~\bibnamefont {Lebrun-Grandié}}, \bibinfo {author} {\bibfnamefont {D.}~\bibnamefont {Arndt}}, \bibinfo {author} {\bibfnamefont {J.}~\bibnamefont {Ciesko}}, \bibinfo {author} {\bibfnamefont {V.}~\bibnamefont {Dang}}, \bibinfo {author} {\bibfnamefont {N.}~\bibnamefont {Ellingwood}}, \bibinfo {author} {\bibfnamefont {R.}~\bibnamefont {Gayatri}}, \bibinfo {author} {\bibfnamefont {E.}~\bibnamefont {Harvey}}, \bibinfo {author} {\bibfnamefont {D.~S.}\ \bibnamefont {Hollman}}, \bibinfo {author} {\bibfnamefont {D.}~\bibnamefont {Ibanez}}, \bibinfo {author} {\bibfnamefont {N.}~\bibnamefont {Liber}}, \bibinfo {author} {\bibfnamefont {J.}~\bibnamefont {Madsen}}, \bibinfo {author} {\bibfnamefont {J.}~\bibnamefont {Miles}}, \bibinfo {author} {\bibfnamefont {D.}~\bibnamefont {Poliakoff}}, \bibinfo {author} {\bibfnamefont {A.}~\bibnamefont {Powell}}, \bibinfo {author} {\bibfnamefont {S.}~\bibnamefont
  {Rajamanickam}}, \bibinfo {author} {\bibfnamefont {M.}~\bibnamefont {Simberg}}, \bibinfo {author} {\bibfnamefont {D.}~\bibnamefont {Sunderland}}, \bibinfo {author} {\bibfnamefont {B.}~\bibnamefont {Turcksin}},\ and\ \bibinfo {author} {\bibfnamefont {J.}~\bibnamefont {Wilke}},\ }\href {https://doi.org/10.1109/TPDS.2021.3097283} {\bibfield  {journal} {\bibinfo  {journal} {IEEE Transactions on Parallel and Distributed Systems}\ }\textbf {\bibinfo {volume} {33}},\ \bibinfo {pages} {805} (\bibinfo {year} {2022})}\BibitemShut {NoStop}%
\bibitem [{\citenamefont {{Turk}}\ \emph {et~al.}(2011)\citenamefont {{Turk}}, \citenamefont {{Smith}}, \citenamefont {{Oishi}}, \citenamefont {{Skory}}, \citenamefont {{Skillman}}, \citenamefont {{Abel}},\ and\ \citenamefont {{Norman}}}]{2011ApJS..192....9T}%
  \BibitemOpen
  \bibfield  {author} {\bibinfo {author} {\bibfnamefont {M.~J.}\ \bibnamefont {{Turk}}}, \bibinfo {author} {\bibfnamefont {B.~D.}\ \bibnamefont {{Smith}}}, \bibinfo {author} {\bibfnamefont {J.~S.}\ \bibnamefont {{Oishi}}}, \bibinfo {author} {\bibfnamefont {S.}~\bibnamefont {{Skory}}}, \bibinfo {author} {\bibfnamefont {S.~W.}\ \bibnamefont {{Skillman}}}, \bibinfo {author} {\bibfnamefont {T.}~\bibnamefont {{Abel}}},\ and\ \bibinfo {author} {\bibfnamefont {M.~L.}\ \bibnamefont {{Norman}}},\ }\href {https://doi.org/10.1088/0067-0049/192/1/9} {\bibfield  {journal} {\bibinfo  {journal} {ApJS}\ }\textbf {\bibinfo {volume} {192}},\ \bibinfo {eid} {9} (\bibinfo {year} {2011})},\ \Eprint {https://arxiv.org/abs/1011.3514} {arXiv:1011.3514 [astro-ph.IM]} \BibitemShut {NoStop}%
\bibitem [{\citenamefont {Harris}\ \emph {et~al.}(2020)\citenamefont {Harris}, \citenamefont {Millman}, \citenamefont {van~der Walt}, \citenamefont {Gommers}, \citenamefont {Virtanen}, \citenamefont {Cournapeau}, \citenamefont {Wieser}, \citenamefont {Taylor}, \citenamefont {Berg}, \citenamefont {Smith}, \citenamefont {Kern}, \citenamefont {Picus}, \citenamefont {Hoyer}, \citenamefont {van Kerkwijk}, \citenamefont {Brett}, \citenamefont {Haldane}, \citenamefont {del R{\'{i}}o}, \citenamefont {Wiebe}, \citenamefont {Peterson}, \citenamefont {G{\'{e}}rard-Marchant}, \citenamefont {Sheppard}, \citenamefont {Reddy}, \citenamefont {Weckesser}, \citenamefont {Abbasi}, \citenamefont {Gohlke},\ and\ \citenamefont {Oliphant}}]{harris2020array}%
  \BibitemOpen
  \bibfield  {author} {\bibinfo {author} {\bibfnamefont {C.~R.}\ \bibnamefont {Harris}}, \bibinfo {author} {\bibfnamefont {K.~J.}\ \bibnamefont {Millman}}, \bibinfo {author} {\bibfnamefont {S.~J.}\ \bibnamefont {van~der Walt}}, \bibinfo {author} {\bibfnamefont {R.}~\bibnamefont {Gommers}}, \bibinfo {author} {\bibfnamefont {P.}~\bibnamefont {Virtanen}}, \bibinfo {author} {\bibfnamefont {D.}~\bibnamefont {Cournapeau}}, \bibinfo {author} {\bibfnamefont {E.}~\bibnamefont {Wieser}}, \bibinfo {author} {\bibfnamefont {J.}~\bibnamefont {Taylor}}, \bibinfo {author} {\bibfnamefont {S.}~\bibnamefont {Berg}}, \bibinfo {author} {\bibfnamefont {N.~J.}\ \bibnamefont {Smith}}, \bibinfo {author} {\bibfnamefont {R.}~\bibnamefont {Kern}}, \bibinfo {author} {\bibfnamefont {M.}~\bibnamefont {Picus}}, \bibinfo {author} {\bibfnamefont {S.}~\bibnamefont {Hoyer}}, \bibinfo {author} {\bibfnamefont {M.~H.}\ \bibnamefont {van Kerkwijk}}, \bibinfo {author} {\bibfnamefont {M.}~\bibnamefont {Brett}}, \bibinfo {author} {\bibfnamefont
  {A.}~\bibnamefont {Haldane}}, \bibinfo {author} {\bibfnamefont {J.~F.}\ \bibnamefont {del R{\'{i}}o}}, \bibinfo {author} {\bibfnamefont {M.}~\bibnamefont {Wiebe}}, \bibinfo {author} {\bibfnamefont {P.}~\bibnamefont {Peterson}}, \bibinfo {author} {\bibfnamefont {P.}~\bibnamefont {G{\'{e}}rard-Marchant}}, \bibinfo {author} {\bibfnamefont {K.}~\bibnamefont {Sheppard}}, \bibinfo {author} {\bibfnamefont {T.}~\bibnamefont {Reddy}}, \bibinfo {author} {\bibfnamefont {W.}~\bibnamefont {Weckesser}}, \bibinfo {author} {\bibfnamefont {H.}~\bibnamefont {Abbasi}}, \bibinfo {author} {\bibfnamefont {C.}~\bibnamefont {Gohlke}},\ and\ \bibinfo {author} {\bibfnamefont {T.~E.}\ \bibnamefont {Oliphant}},\ }\href {https://doi.org/10.1038/s41586-020-2649-2} {\bibfield  {journal} {\bibinfo  {journal} {Nature}\ }\textbf {\bibinfo {volume} {585}},\ \bibinfo {pages} {357} (\bibinfo {year} {2020})}\BibitemShut {NoStop}%
\bibitem [{\citenamefont {Virtanen}\ \emph {et~al.}(2020)\citenamefont {Virtanen}, \citenamefont {Gommers}, \citenamefont {Oliphant}, \citenamefont {Haberland}, \citenamefont {Reddy}, \citenamefont {Cournapeau}, \citenamefont {Burovski}, \citenamefont {Peterson}, \citenamefont {Weckesser}, \citenamefont {Bright}, \citenamefont {{van der Walt}}, \citenamefont {Brett}, \citenamefont {Wilson}, \citenamefont {Millman}, \citenamefont {Mayorov}, \citenamefont {Nelson}, \citenamefont {Jones}, \citenamefont {Kern}, \citenamefont {Larson}, \citenamefont {Carey}, \citenamefont {Polat}, \citenamefont {Feng}, \citenamefont {Moore}, \citenamefont {{VanderPlas}}, \citenamefont {Laxalde}, \citenamefont {Perktold}, \citenamefont {Cimrman}, \citenamefont {Henriksen}, \citenamefont {Quintero}, \citenamefont {Harris}, \citenamefont {Archibald}, \citenamefont {Ribeiro}, \citenamefont {Pedregosa}, \citenamefont {{van Mulbregt}},\ and\ \citenamefont {{SciPy 1.0 Contributors}}}]{2020SciPy-NMeth}%
  \BibitemOpen
  \bibfield  {author} {\bibinfo {author} {\bibfnamefont {P.}~\bibnamefont {Virtanen}}, \bibinfo {author} {\bibfnamefont {R.}~\bibnamefont {Gommers}}, \bibinfo {author} {\bibfnamefont {T.~E.}\ \bibnamefont {Oliphant}}, \bibinfo {author} {\bibfnamefont {M.}~\bibnamefont {Haberland}}, \bibinfo {author} {\bibfnamefont {T.}~\bibnamefont {Reddy}}, \bibinfo {author} {\bibfnamefont {D.}~\bibnamefont {Cournapeau}}, \bibinfo {author} {\bibfnamefont {E.}~\bibnamefont {Burovski}}, \bibinfo {author} {\bibfnamefont {P.}~\bibnamefont {Peterson}}, \bibinfo {author} {\bibfnamefont {W.}~\bibnamefont {Weckesser}}, \bibinfo {author} {\bibfnamefont {J.}~\bibnamefont {Bright}}, \bibinfo {author} {\bibfnamefont {S.~J.}\ \bibnamefont {{van der Walt}}}, \bibinfo {author} {\bibfnamefont {M.}~\bibnamefont {Brett}}, \bibinfo {author} {\bibfnamefont {J.}~\bibnamefont {Wilson}}, \bibinfo {author} {\bibfnamefont {K.~J.}\ \bibnamefont {Millman}}, \bibinfo {author} {\bibfnamefont {N.}~\bibnamefont {Mayorov}}, \bibinfo {author} {\bibfnamefont
  {A.~R.~J.}\ \bibnamefont {Nelson}}, \bibinfo {author} {\bibfnamefont {E.}~\bibnamefont {Jones}}, \bibinfo {author} {\bibfnamefont {R.}~\bibnamefont {Kern}}, \bibinfo {author} {\bibfnamefont {E.}~\bibnamefont {Larson}}, \bibinfo {author} {\bibfnamefont {C.~J.}\ \bibnamefont {Carey}}, \bibinfo {author} {\bibfnamefont {{\.I}.}~\bibnamefont {Polat}}, \bibinfo {author} {\bibfnamefont {Y.}~\bibnamefont {Feng}}, \bibinfo {author} {\bibfnamefont {E.~W.}\ \bibnamefont {Moore}}, \bibinfo {author} {\bibfnamefont {J.}~\bibnamefont {{VanderPlas}}}, \bibinfo {author} {\bibfnamefont {D.}~\bibnamefont {Laxalde}}, \bibinfo {author} {\bibfnamefont {J.}~\bibnamefont {Perktold}}, \bibinfo {author} {\bibfnamefont {R.}~\bibnamefont {Cimrman}}, \bibinfo {author} {\bibfnamefont {I.}~\bibnamefont {Henriksen}}, \bibinfo {author} {\bibfnamefont {E.~A.}\ \bibnamefont {Quintero}}, \bibinfo {author} {\bibfnamefont {C.~R.}\ \bibnamefont {Harris}}, \bibinfo {author} {\bibfnamefont {A.~M.}\ \bibnamefont {Archibald}}, \bibinfo {author}
  {\bibfnamefont {A.~H.}\ \bibnamefont {Ribeiro}}, \bibinfo {author} {\bibfnamefont {F.}~\bibnamefont {Pedregosa}}, \bibinfo {author} {\bibfnamefont {P.}~\bibnamefont {{van Mulbregt}}},\ and\ \bibinfo {author} {\bibnamefont {{SciPy 1.0 Contributors}}},\ }\href {https://doi.org/10.1038/s41592-019-0686-2} {\bibfield  {journal} {\bibinfo  {journal} {Nature Methods}\ }\textbf {\bibinfo {volume} {17}},\ \bibinfo {pages} {261} (\bibinfo {year} {2020})}\BibitemShut {NoStop}%
\bibitem [{\citenamefont {Hunter}(2007)}]{Hunter:2007}%
  \BibitemOpen
  \bibfield  {author} {\bibinfo {author} {\bibfnamefont {J.~D.}\ \bibnamefont {Hunter}},\ }\href {https://doi.org/10.1109/MCSE.2007.55} {\bibfield  {journal} {\bibinfo  {journal} {Computing in Science \& Engineering}\ }\textbf {\bibinfo {volume} {9}},\ \bibinfo {pages} {90} (\bibinfo {year} {2007})}\BibitemShut {NoStop}%
\bibitem [{\citenamefont {Dalcin}\ \emph {et~al.}(2019)\citenamefont {Dalcin}, \citenamefont {Mortensen},\ and\ \citenamefont {Keyes}}]{mpi4py-fft}%
  \BibitemOpen
  \bibfield  {author} {\bibinfo {author} {\bibfnamefont {L.}~\bibnamefont {Dalcin}}, \bibinfo {author} {\bibfnamefont {M.}~\bibnamefont {Mortensen}},\ and\ \bibinfo {author} {\bibfnamefont {D.~E.}\ \bibnamefont {Keyes}},\ }\href {https://doi.org/https://doi.org/10.1016/j.jpdc.2019.02.006} {\bibfield  {journal} {\bibinfo  {journal} {Journal of Parallel and Distributed Computing}\ }\textbf {\bibinfo {volume} {128}},\ \bibinfo {pages} {137 } (\bibinfo {year} {2019})}\BibitemShut {NoStop}%
\end{thebibliography}%

\end{document}